%% file: messenger05_arxiv.tex
\documentclass[onefignum,onetabnum]{siamart190516}

\input{ex_shared}

\begin{document}

\maketitle

\begin{abstract}
Interacting particle system (IPS) models have proven to be highly successful for describing the spatial movement of organisms. However, it has proven challenging to infer the interaction rules directly from data. In the field of equation discovery, the Weak form Sparse Identification of Nonlinear Dynamics (WSINDy) methodology has been shown to be very computationally efficient for identifying the governing equations of complex systems, even in the presence of substantial noise.  Motivated by the success of IPS models to describe the spatial movement of organisms, we develop WSINDy for second order IPSs to model the movement of communities of cells.  Specifically, our approach learns the  directional interaction rules that govern the dynamics of a heterogeneous population of migrating cells. Rather than aggregating cellular trajectory data into a single best-fit model, we learn the models for each individual cell. These models can then be efficiently classified according to the active classes of interactions present in the model. From these classifications, aggregated models are constructed hierarchically to simultaneously identify different \textit{species} of cells present in the population and determine best-fit models for each species. We demonstrate the efficiency and proficiency of the method on several test scenarios, motivated by common cell migration experiments.  
\end{abstract}


\begin{multicols}{2}

\section{Introduction}


Systems of autonomous agents are ubiquitous in the natural world. Research into their behavior has led to a plethora of proposed mathematical models, including the agent-based ``boids'' model \cite{reynolds1987flocks}, ordinary differential equation models for milling and flocking \cite{levine2000self,cucker2007emergent}, and nonlocal partial differential equations \cite{mogilner1999non,toner1998flocks}, to name a few. A general framework for rigorous analysis of these models is by now very mature \cite{carrillo2021mean}. 

Identifying the rules of interaction between agents is necessary for predicting and influencing the cooperative abilities of any such system, whether composed of autonomous robots, large multi-cellular animals, single-celled organisms, or even molecules. Methods for inferring the rules of interaction between agents using observed trajectory data have continued to advance since the early 2000's. Several of the principled techniques include force-matching \cite{eriksson2010determining,katz2011inferring}, linear regression \cite{lukeman2010inferring,lu2020learning}, mean-field formulations \cite{supekar2021learning,messenger2021learning}, information-theoretic tools \cite{pilkiewicz2020decoding}, underdamped Langevin regression \cite{bruckner2021learning,bruckner2020inferring}, Gaussian processes \cite{feng2021data}, and even a method based on topological rather than metric distances \cite{BalleriniEtAl2008pnas}.

These and related techniques have been successfully used to identify the dominant drivers of collective behavior in a variety of social and biological systems \cite{SchaerfHerbert-ReadWard2021JRSocInterface, SumpterSzorkovszkyKotrschalEtAl2018PhilTransRSocB, Akaike1974IEEETransAutomControl}, including schools of fish \cite{WardSchaerfHerbert-ReadEtAl2017RSocopensci,TunstromKatzIoannouEtAl2013PLoSComputBiol}, flocks of birds \cite{chen2017anisotropic,paranjape2018robotic}, and pedestrian traffic \cite{Warren2018CurrDirPsycholSci}, all directly incorporating measured trajectory data. While popular methods, such as force-matching, are useful in identifying fields of vision and spatial statistics of interactions, they cannot easily disentangle the combined effects of multiple forces (e.g.\ attraction, repulsion, and alignment) \cite{mudaliar2020examination,escobedo2020data}, let alone different interactions between multiple species of neighbors. This limits the classes of models they can identify and implies that new methods must be developed for heterogeneous populations. 
 
The field of {\it equation discovery} is a highly active area of research \cite{LagergrenNardiniMichaelLavigneEtAl2020ProcRSocA,nardini2021learning,schmidt2009distilling,rudy2017data,brunton2016discovering,schaeffer2017learning,udrescu2020ai,long2018pde,champion2019data} as it offers tools to directly learn governing differential equations.  This approach is not only useful in prediction and validation, but can be used to simultaneously identify multiple active modes of inter-agent communication, such as repulsion, velocity alignment, and attraction. In this work, we tackle the problem of identifying governing equations for an interacting particle system (IPS) with multiple interacting species. Our proposed approach is completely naive with regard to species membership in order to specifically address problems of heterogeneity in collective cell migration \cite{tang2012understanding}. 

Motivated by existing hypotheses regarding the anistropy of cell-cell interactions \cite{sarkar2021minimal,holmes2017mathematical,notbohm2016cellular}, we introduce our framework in the context of {\it directional interaction} models, as defined below. Moreover, we note that the documented significance of anisotropic interactions in general collective systems \cite{eftimie2007complex,BernardiEftimiePainter2021BullMathBiol,zmurchok2018direction} suggests that our approach may have wide applicability.
 
\subsection{Heterogeneous Populations} 
 
Many collective populations arising in nature are inherently heterogeneous, with the rules of interaction varying across different subsets of the population. This is readily observable in complex mammalian populations, but is also seen in simpler organisms, such as honey bee swarms, where bees divide into scout and worker bee roles \cite{seeley2006group}. The advantages of heterogeneity in collective behavior have even inspired search optimization algorithms \cite{engelbrecht2010heterogeneous,kengyel2015potential}.

At the level of microorganisms, cells have been observed to adopt leader-like and follower-like roles during migration \cite{vishwakarma2018mechanical}, without the aid of a central nervous system. Individual cell speed and persistence of motion have also been determined to be functions of the age and size of the cell \cite{lan2021decomposition,bonneton1999relationship,boehm2001cell}, which may lead to heterogeneous responses to stimuli from neighboring cells. The mechanisms which produce these heterogeneities, and the extent to which heterogeneity is present in a given cell population, are current subjects of debate \cite{nardini2016modeling,SchumacherMainiBaker2017CellSystems, HaegerWolfZegersEtAl2015TrendsCellBiol}. Data-driven techniques may be useful in formulating accurate mathematical models in the presence of heterogeneity. 

The authors of \cite{zhong2020data} develop a highly versatile method for inferring explicit rules of interaction in a heterogeneous population, although it is assumed that species membership is known {\it a priori}. Several recent works have offered methods of assessing the degree of population heterogeneity \cite{SchumacherMainiBaker2017CellSystems,SchaerfHerbert-ReadWard2021JRSocInterface}, yet these methods do not provide explicit mathematical models for the different populations. In contrast, the method presented here allows one to classify the given population into different species according to the heterogeneous interaction rules present, and produces explicit mathematical models for each species as a byproduct. 

 \subsection{Directional Interaction Forces}
 
 It is now well-known that simple radial interaction models are incapable of explaining many observed collective behaviors in biological settings, and that directionally-dependent interaction rules, based on a limited field of view or sensing angle, offer a significant advantage \cite{eftimie2007complex,chen2017anisotropic,cai2006modelling,carrillo2010particle,evers2015anisotropic}. At the cellular level, directional dependence of cell-cell interaction has been proposed in the context of intracellular polarization \cite{holmes2017mathematical}, however the cellular sensing range is not immediately obvious, since a migrating cell does not have an obvious ``field of view''. Recent works have sought to quantify the degree to which interactions are density-dependent \cite{browning2020identifying}, but not which directional modes (radial, dipolar, quadrupolar, etc.) are dominant during a collective migration event. 
 
 In addition to providing an explanation for certain observed phenomena \cite{zmurchok2018direction}, directional interaction rules are capable of generating {\it spontaneous migration}, due to the total directional force between particles not being conserved in general. In the modeling of active matter systems (such as migrating cells) \cite{ramaswamy2010mechanics,supekar2021learning}, such symmetry breaking is commonly generated by a combination of Brownian forcing and a self-propulsion device \cite{vicsek2012collective}. However, it is not clear that self-propulsion is an appropriate mechanism for modeling cellular movement (in comparison to fish, which are constantly swimming). Directional forces may then be an important mechanism for symmetry breaking and spontaneous cellular migration.

\subsection{Weak-Form Sparse Identification of Nonlinear Dynamics}

At its core, our method involves learning ordinary differential equations for cells using available trajectory data. For this we employ the weak-form sparse identification of nonlinear dynamics algorithm (WSINDy), which has been shown to successfully identify governing equations from data at the levels of ordinary different equations \cite{messenger2020weak}, partial differential equations \cite{messenger2020weakpde}, 1st-order interacting particle systems \cite{messenger2021learning}, and even works in a small-storage streaming scenario \cite{MessengerDallAneseBortz2022arXiv220303979}.

A significant advantage of the WSINDy method is that it identifies a single governing equation which can be analyzed and simulated using conventional techniques of applied mathematics. It does not involve any black-box algorithms or mappings as would be generated in using a neural network-based approach.\footnote{See  \cite{Bhat2020MachineLearningandKnowledgeDiscoveryinDatabases} for an example where the authors first learn a neural network model of the potential and then uses sparse identification to learn the algebraic form of the potential.}

Several alternative methods have been developed to accomplish the equation learning task for particle systems. In particular, Lu et al. \cite{lu2020learning} develop a method for learning general feature-dependent 2nd-order interaction rules for heterogeneous populations, where features may include directional interaction forces, speed dependence, and so on. The differences between this and our work are the following. (i) We are performing the {\it unsupervised} learning task of classifying agents by their interaction rules, whereas Lu et al.'s work assumes knowledge of the species membership, (ii) we are interested in {\it sparse model representations}, in particular selection of the correct modes of interaction (e.g.\ attractive, repulsive, alignment, and drag force), whereas Lu's work assumes knowledge of both the feature-dependence and types of forces present (e.g.\ for planetary systems, priori knowledge is used to rule out the presence of an alignment force). (iii) Lastly, models are initially extracted from {\it single-cell trajectories}. As described in the next section, rather than aggregating {\it data} which may come from multiple cell species, we aggregate {\it models} which are likely to describe the same species, and then use the aggregate model to perform classification.

\subsection{Single-Cell Learning and Model Clustering}

With a possibly heterogeneous population of cell trajectories available, one is tasked with the problem of deciding how to aggregate the data. If knowledge of the underlying species membership is available, a more accurate model can be inferred by pooling data from all individuals of a given species. On the other hand, pooling data from multiple species into a single model can result in a highly {\it inaccurate} model depending on the difference between interaction rules in each species. In general, there exists a spectrum of possible pooling strategies, ranging from learning {\it few models} from {\it large subsets} of the population, to learning {\it many models} from {\it small subsets} of the populations. The former intrinsically produces models with high bias and low variance, while the latter produces models with low bias and high variance. Such pooling strategies have been recently explored in \cite{FaselKutzBruntonEtAl2022ProcRSocMathPhysEngSci}, where it is found that identifying a single model can be improved by pooling models learned from subsets of the data. However, this has not been extended to classifying the data itself into species, and finding a model for each species. Moreover, the IPS setting offers a particular advantage on the subject of model validation, as data can easily be assimilated into forward simulations.

In this work we investigate the extreme case of learning an individual model $\CalM_i$ for the $i$-th individual trajectory, and then clustering the set of learned models $\CalM:= \{\CalM_1,\dots,\CalM_N\}$ according to their identified modes of interaction.  This approach is counter-intuitive because there is no guarantee that a single cell trajectory will provide enough information on the interaction rules of its species. To be able to classify cells using the (potentially) insufficiently informative trajectories, we developed an ad hoc recursive classifier which we show (in Section \ref{sec:results}) accurately clusters and sorts the models into species. This approach prevents any contamination that may result from combining trajectories of multiple species. 

Once the models are clustered, an aggregate model $\overline{\CalM}$ is computed by averaging the models in $\CalM$ belonging to the most populous cluster.  The model $\overline{\CalM}$ is then used to classify cells via forward simulations which are made highly efficient by directly incorporating the data. In particular, for each trajectory in the dataset, we use $\overline{\CalM}$ to simulate a new trajectory, but with all neighbor interactions computed using the data. That is, only the new trajectory is propagated forward in time by model $\overline{\CalM}$, while the rest of the population is simply the data itself. This can then be trivially parallelized, reducing an $\CalO(N^2)$ computational cost per timestep to $N$ cores performing $\CalO(N)$ updates per timestep with no communication overhead.

We show through examples below that this hierarchical model-pooling and validation procedure produces both correct species classification and accurate governing equations, despite individual cell trajectory data carrying low levels of information. For further information on the classification algorithm, see Section \ref{sec:alg}.

\subsection{Paper outline}

In Section \ref{sec:DIPM} we discuss the general form of directional interacting particle models that will be assumed in the learning process. In Section \ref{sec:alg} we introduce our model selection and classification algorithm, which is composed of the five steps: (a) learn single-cell models, (b) cluster learned models according to force modes, (c) form an aggregate model by averaging models in the largest cluster, (d) validate the aggregate model using data-driven forward simulations, (e) classify cells according to performance under the aggregate model. In Section \ref{sec:results} we examine the performance of the algorithm in learning and classifying homogeneous and heterogeneous populations of one, two, and three species. We discuss possible next directions in Section \ref{sec:discussion}. Some additional information and a summary of notation are included in the appendix. 

\section{Directional Interacting Particle Models}\label{sec:DIPM}

\begin{figure*}
\input{tikz_theta}
\caption{Diagram of social interactions depending on angle $\theta_{ij}$ between cell $i$'s velocity and cell $j$'s position relative to $i$.}\label{thetapic}
\end{figure*}
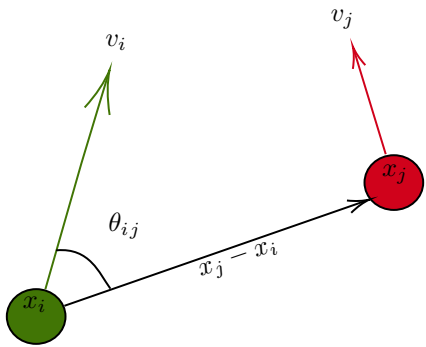

We use a general 2nd-order directional interaction model framework, where the position and velocity $(x_i,v_i)$ of cell $i$ are governed by the differential equations
\begin{equation}\label{dfm}
\begin{dcases}
\ddot{x}_i = \frac{1}{N_\text{tot}}\sum_{j=1}^{N_\text{tot}} f_\text{a-r}(|x_i-x_j|,\theta_{ij})(x_i-x_j)\\
\qquad +\frac{1}{N_\text{tot}}\sum_{j=1}^{N_\text{tot}} f_\text{align}(|x_i-x_j|,\theta_{ij})(v_i-v_j)\\ \qquad+\frac{1}{N_\text{tot}}\sum_{j=1}^{N_\text{tot}} f_\text{drag}(|v_i|,\theta_{ij})v_i+ \eta_i.
\end{dcases}
\end{equation}
Here, $\theta_{ij}$ is the angle between $v_i$ and $ x_j-x_i$ (see the diagram in Figure \ref{thetapic}) and $(\eta_i)_{i\in[N_{tot}]}$ are white noise processes.  The attractive-repulsive force $f_{\text{a-r}}$, alignment force $f_\text{align}$, and the drag force $f_\text{drag}$ define the rules by which cell $i$ communicates with the rest of the population. Our primary objective is to identify a set of interaction rules $\{(f_\text{a-r}, f_\text{align}, f_\text{drag})_\ell\}_{1\leq \ell \leq S}$, one for each of the $S$ species present in the population.

\subsection{Directionality $\theta_{ij}$}

As mentioned above, directional variation in the interaction forces between cells 
can arise from various factors, including intracellular polarization, or heterogeneous distribution of membrane-bound receptors, asymmetry in the protrusion/contraction of lamellopodia as the cell crawls on the substrate, and so on. In the current study, we assume that each of these effects is unobservable, hence we model the aggregate directional effect using the angles $\theta_{ij}$, depicted in Figure \ref{thetapic}. Dependence on angle $\theta_{ij}$ allows for interactions between cell $i$ and cell $j$ to vary depending on the direction of motion\footnote{Note that asymmetric interactions $\theta_{ij}\neq \theta_{ji}$ lead to symmetry breaking and spontaneous cell migration from an initially motionless state.}. Put another way, in the reference frame of cell $i$, the polar coordinates $r_{ij} = |x_i-x_j|$ and $\theta_{ij}$ allow one to represent any interaction force that varies over the 2D plane. 

In this study we restrict the angular dependence to $\{1, \cos(\theta_{ij}), \cos(2\theta_{ij})\}$, which allows for a combination of radial, dipolar, and quadrupolar interactions (see Figure \ref{artforce} for examples of dipolar (right) and quadrupolar (left) forces used in this study). Higher-order directionality can usually be assumed to be negligible, however extension to higher modes is straightforward.

\subsection{Attractive-repulsive force $f_\text{a-r}$ }

The interaction force $f_\text{a-r}$ acts along the vector from cell $i$ to cell $j$ and captures short-range repulsion and long-range signaling. Many IPS models include only an attractive-repulsive force, due to its extensive pattern-forming capabilities \cite{topaz2004swarming,fetecau2011swarm}.

We impose the following natural constraints on $f_\text{a-r}$:

\begin{equation}\label{fconst}
f_\text{a-r}(r,\theta) \begin{dcases} 
\geq 0, & 0\leq r< r_{nf} \\ 
\leq 0, & r\geq r_{ff} \\ 
\in \text{span}\{1,\cos(\theta), \cos(2\theta)\}, & \text{every $r$ fixed,}
\end{dcases}
\end{equation}
where $r_{nf}$ is the {\it near-field threshold} and $r_{ff}$ is the {\it far-field threshold}, i.e., a large distance.

The first inequality enforces that $f_\text{a-r}$ is near-field repulsive, which must be true by volume exclusion. In practice we define the  $r_{nf}$ implicitly by \[\Pbb\left(|x_i-x_j|<r_{nf}\right)=p_{nf},\]
where in this work we set $p_{nf}=0.001$, and the dataset is used to compute the probability. This states that the near-field region is defined by short-distance interactions which account for less than $0.1\%$ of observed neighbor-neighbor distances.

The second equality enforces long-range decay, as well as model stability. Decay is natural since interactions can be expected to be small outside of some large distance $r_{ff}$. We enforce that interactions are {\it attractive} at large distances (allowing for decay as well), so that in simulation the particles do not spread to infinity. We set $r_{ff}=1$ throughout, although $r_{ff}$ can easily be chosen from the data (e.g.\ $r_{ff} = 50r_{nf}$ corresponds to an effective interaction range of 50 cell radii).  (See Appendix \ref{app:hp} for resulting values of $r_{nf}$ and $r_{ff}$ and other hyperparameters for examples below). 

The third set inclusion simply reiterates the assumptions on directionality described above.

\subsection{Alignment force $f_\text{align}$}

The alignment force $f_\text{align}$ captures cells' tendency to match the velocity of neighboring cells. There are many theories as to how this arises physically \cite{tambe2011collective,sarkar2021minimal}. Perhaps protrusions from cells inform the cell about the bulk direction of motion, which would be a very local effect. However, alignment models which have been proven to lead to flocking depend on sufficiently {\it long-range} alignment \cite{cucker2007emergent}.

We impose the following constraints on $f_\text{align}$:

\begin{equation}\label{hconst}
f_\text{align}(r,\theta) \begin{dcases} 
\leq 0, & 0\leq r \\ 
\in \text{span}\{1,\cos(\theta), \cos(2\theta)\}, & \text{every $r$ fixed}
\end{dcases}
\end{equation}

The first inequality enforces that $f_\text{align}$ is non-positive, which is necessary for the constant velocity state $v_i=v_j=v$ to be a stable configuration. If not, small perturbations away from $v_i=v_j$ result in cells {\it accelerating} away from each other, which is a redundant force given that cells can be pushed away from each other through $f_\text{a-r}$ (it is also not hard to see that $f_\text{align}> 0$ is unphysical). The second constraint restricts the alignment force to be a combination of radial, dipolar, or quadrupolar modes, similar to $f_\text{a-r}$.

\subsection{Drag force $f_\text{drag}$}

The drag force $f_\text{drag}$ captures energy expenditure due to general  resistance to motion (resulting e.g. from substrate roughness), however we allow an angular dependence on $\theta_{ij}$ to capture possible decreases or increases in drag depending on local neighbor distribution. For this we impose the following constraints:

\begin{equation}\label{dconst}
f_\text{drag}(s,\theta) \begin{dcases} 
\leq 0, & 0\leq s < \infty \\ 
\in \text{span}\{1,\cos(\theta)\}, & \text{every $s$ fixed}
\end{dcases}
\end{equation}
where $s$ indicates the speed of the cell. The force $f_\text{drag}$ is chosen to be negative so that cells do not have a ``self-propulsion'' device. As mentioned previously, many models of active matter include self-propulsion as a partial mechanism for symmetry breaking and general non-equilibrium effects. To reiterate, we do not expect cells to have a self-propulsion device, in fact, we wish to learn how migration occurs spontaneously, incited by pairwise interactions. In addition, a positive drag force leads to populations spreading outside of the range of meaningful interactions. In this way, negative drag is computationally beneficial, as it leads to improved model stability.

\section{Algorithm}\label{sec:alg}

\begin{figure*}
		\includegraphics[trim={0 0 0 0},clip,width=1\textwidth]{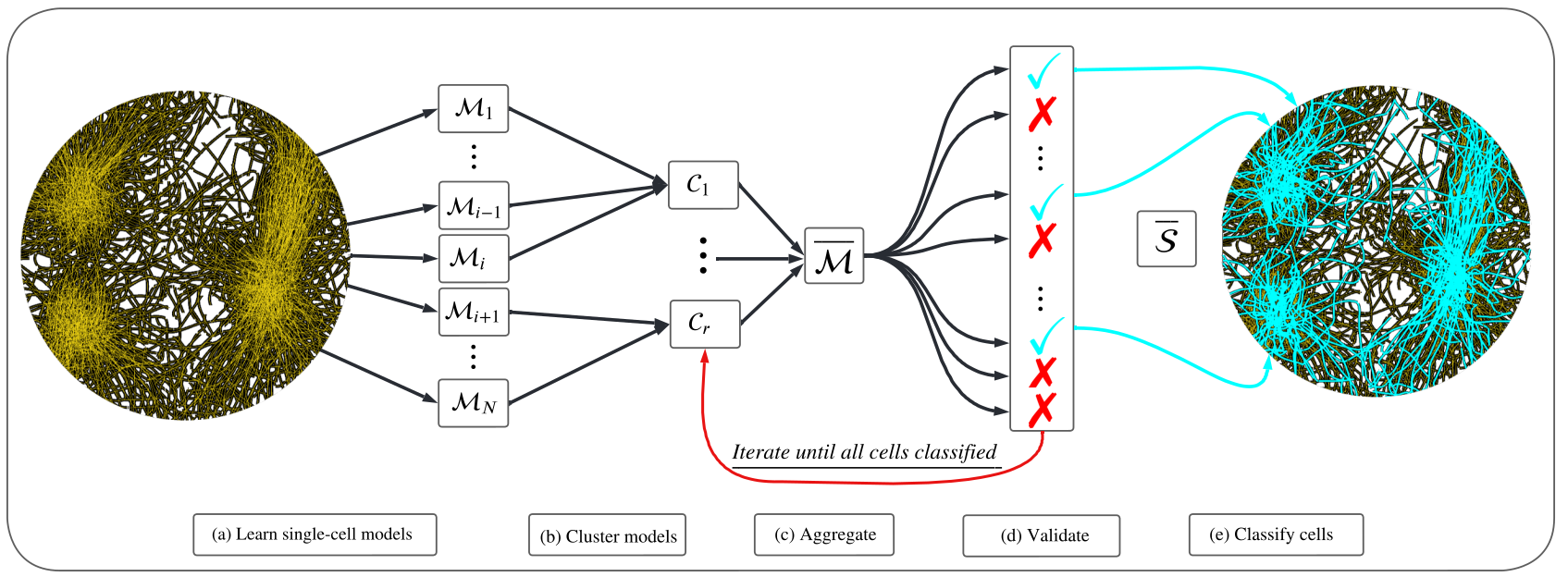} 
		\caption{Classification pipeline for cells from heterogeneous populations. (a) An ensemble of models $\CalM= \{\CalM_1,\dots,\CalM_N\}$ is learned, each from an individual trajectory; (b) $\CalM$ is partitioned into clusters $\CalC = \{\CalC_1,\dots,\CalC_r\}$ according to active forces in each model; (c) models in the largest cluster $\overline{\CalC}$ are averaged together, giving $\overline{\CalM}$; (d) $\overline{\CalM}$ is validated along each individual trajectory; (e) validation errors are classified, producing an identified species $\overline{\CalS}$ (cyan checkmarks) and the remaining cells (red X's) are returned to step (c) to be clustered again. Steps (b)-(e) repeat until all cells are classified. Note that the number and members of model clusters $\CalC$ and resulting aggregate model $\overline{\CalM}$ will change each iteration depending on the identity of remaining unlabeled cells.}     
		\label{fig:algo}
\end{figure*}


Our algorithm involves first learning an ensemble of directional force models $\CalM = \{\CalM_1,\dots,\CalM_N\}$, that is, one model for each of the $N$ {\it focal cells} selected for learning. We next group models into clusters $\CalC := \{\CalC_1,\dots,\CalC_r\}$ and compute an aggregrate model $\overline{\CalM}$ from the largest cluster, denoted by $\overline{\CalC}$. A new species $\overline{\CalS}$ is then identified, with membership in $\overline{\CalS}$ determined by the accuracy of data-driven forward simulations of model $\overline{\CalM}$. Cells in the species $\overline{\CalS}$ are then removed from the population and the remaining cells are returned to the clustering phase. More explicitly, the algorithm is composed of the following steps, which are visualized in Figure \ref{fig:algo}. 

\begin{enumerate}[label=(\alph*)]
\item Identify individual cell models $\CalM = \{\CalM_1,\dots,\CalM_N\}$ using the WSINDy algorithm 
\item Partition learned models $\CalM$ into clusters $\{\CalC_1,\dots,\CalC_r\}$ according to active force modes
\item  Identify the largest cluster $\overline{\CalC}$ and average together the models in $\overline{\CalC}$ to arrive at a single model $\overline{\CalM}$
\item Simulate the model $\overline{\CalM}$ separately for each remaining unlabeled cell, using the data to calculate neighbor interactions 
\item Classify cells based on simulation error under  $\overline{\CalM}$ and label the lowest-error class as the new species $\overline{\CalS}$
\item Remove cells in $\overline{\CalS}$ from the population and return to step (b)
\end{enumerate}
To summarize, the result is a set of $S$ models and species $\{(\overline{\CalM}_\ell,\overline{\CalS}_\ell)\}_{\ell=1}^S$, where each model $\overline{\CalM}_\ell$ is constructed from an average of individual models within a cluster. We use the notation\footnote{See Table \ref{table:notation} for a complete list of notations used.} of $\CalM_i$ to be the model for the $i$-th cell, $\CalC_j$ to be the $j$-th cluster of models, and $\overline{\CalS}_{\ell}$ to be the set of cells identified as the $\ell$-th species and which obey model $\overline{\CalM}_{\ell}$.

We now give more detail on steps (a)-(f), including stopping criteria, leaving the more technical aspects to the appendix. 

\subsection{Learning single-cell models}

The first step in the algorithm is to learn an ensemble of {\it single-cell} models, one for each of the $N$ focal cell selected from the $N_\text{tot}$ total cells tracked during the experiment\footnote{For the artificial data used here, the set of focal cells is the entire population ($N=N_{tot}$), however we discuss in the conclusion various situations in which a subset of $N<N_\text{tot}$ focal cells should be chosen.}. By ``single-cell'' model, we mean that the scope of each model is limited to learning only the dynamics of its focal cell, however data from the remaining tracked cells are incorporated to learn the interaction forces.

\subsubsection{WSINDy}

The main ingredient in learning single-cell models in $\CalM$ is the WSINDy algorithm, together with careful choices for the bases used to represent the three main forces $f_\text{a-r}$, $f_\text{align}$, and $f_\text{drag}$. Each cell $i$ is identified by a position and velocity $(x_i(t),v_i(t))$ which we assume is well-approximated by a 2nd-order model of the form \eqref{dfm}. The dynamics take the general form
\begin{equation}\label{3ode}
\ddot{x}_i(t) = F_i(X(t),V(t)),
\end{equation}
where $(X(t),V(t)) \in \mathbb{R}^{2dN_\text{tot}}$ denotes the entire population of positions and velocities in the colony at time $t$.  We then assume that we have available a dataset of positions $\Xbf = (\xbf_1(t_k),\dots,\xbf_{N_\text{tot}}(t_k))_{k=1}^L$ sampled from the system $X$ at $L$ timepoints. Our goal is to identify $F_i$ using $\Xbf$.

The SINDy approach involves representing $F_i$ as a sparse linear combination of basis elements $\Theta(X,V):=(f_j(X,V))_{1\leq j\leq J}$, such that
\[F_{i,d^\prime}(X,V) = \sum_{j=1}^J\wstar_{i,j}f_{j,d^\prime}(X,V),\]
where subscript $d^\prime$ indicates the spatial coordinate ($d^\prime\in\{1,2\}$ in this study). The basis $\Theta$ is chosen by the user and determines the accuracy of the learned model as well as the conditioning of the WSINDy algorithm. 

The available cell position data $\Xbf$ is used to approximate velocities $\Vbf:=\dot{\Xbf}\approx \dot{X}$ and accelerations $\ddot{\Xbf}\approx \ddot{X}$, using e.g.\ finite differences, leading from \eqref{3ode} to the data-driven linear system
\begin{equation}\label{3sindy}
\ddot{\xbf}_i\approx \Theta(\Xbf,\Vbf)\wstar_i.
\end{equation}
With some abuse of notation, we denote by $\Theta(\Xbf,\Vbf)$ the matrix that results from evaluating the basis $\Theta(X,V)$ at the time-series data $(\Xbf,\Vbf)$. The entries are $\Theta(\Xbf,\Vbf)_{k+(d^\prime-1),j} = f_{j,d^\prime}(\Xbf(t_k),\Vbf(t_k))$.

The data $\Xbf$ is often corrupted by measurement noise, which leads to inaccurate computations of derivatives $\Vbf$. For the current setting, the standard SINDy approach just outlined requires {\it second-order} derivatives $\ddot{\Xbf}$, which are even less accurate to compute from noisy data. To prevent some of the corruption from noise, we can use the weak form, which leads to WSINDy. Returning to equation \eqref{3ode}, we convert to the weak form by multiplying by a {\it test function} $\phi(t)$ and integrating in time,
\begin{equation}\label{3intphi}
\lan \phi,\ddot{x}_i\ran:= \lan \phi, F_i(X ,V)\ran.
\end{equation}
where the inner product $\lan \cdot,\cdot \ran$ denotes the time intergral
\[\lan f,g\ran = \int_{-\infty}^\infty f(t)g(t)dt.\]
Choosing $\phi$ to be twice differentiable and zero outside of some interval $(a,b)$, we then integrate by parts twice on the left-hand side to arrive at 
\[\langle\ddot{\phi}, x_i \rangle = \lan \phi, F_i(X,V) \ran,\]
so that the second derivative has been removed from $x_i$ and placed on $\phi$. Choosing a basis of test functions $\Phi:=(\phi_q)_{1\leq q\leq Q}$, we build the weak-form linear system 
\begin{equation}\label{3wsindy}
\bbf^{(i)} \approx \Gbf^{(i)}\what^{(i)}
\end{equation}
where $\bbf^{(i)}_{q + (d^\prime-1)} = \langle \ddot{\phi}_q,\xbf_{i,d^\prime}\rangle$ and $\Gbf^{(i)}_{q + (d^\prime-1),j} = \lan \phi_q, f_{j,d^\prime}(\Xbf,\Vbf)\ran$.

As well as choosing $\Theta$, in order to compute $(\Gbf^{(i)},\bbf^{(i)})$, we need to compute\footnote{In some cases we can use weak-form integration to eliminate this step but we do not pursue this in the current work.} $\Vbf$ from the position data $\Xbf$, choose a test function basis $\Phi$, and discretize integrals appearing in the linear system. For simplicity, we compute $\Vbf$ using 2nd-order centered finite difference, although a number of methods exist for numerical differentiation from data \cite{knowles2014methods,van2020numerical}. For integration, we use the trapezoidal rule, and we use test functions of the form
\begin{equation}\label{3phi}
\phi_q(t) = \max\left(1-\left(\frac{t-t_q}{m\Delta t}\right)^2,\ 0\right)^p,
\end{equation}
for shape parameters $m$ and $p$, and timestamps $t_q$ in the range of the available time series. We refer to \cite{messenger2020weak,messenger2020weakpde} for methods of choosing $(m,p,t_q)$ and for accuracy and robustness results concerning the trapezoidal rule combined with this particular class of test functions. In this work, we use the changepoint algorithm in \cite{messenger2020weakpde} with $\tau = 10^{-10}$ and $\widehat{\tau}=3$, leading to $m\in\{31,\dots,38\}$ and $p\in\{8,9\}$ (Table \ref{tabel:hp} lists these values used in the examples below). Since the time series below are relative short ($L=200$ or $L=400$), we use all available $t_q$, i.e. $(t_q)_{q=1}^Q = (m\Delta t,\dots, (L-m-1)\Delta t)$ so that $Q=L-2m$.

\subsubsection{Regression}

We solve the linear system \eqref{3wsindy} for coefficients $\what^{(i)}$ by approximately solving the following constrained sparse regression problem:
\begin{equation}\label{optprob}
\what^{(i)} = \argmin_{\wbf \ s.t. \ \Cbf\wbf\leq \dbf}\left\{ \nrm{\Gbf^{(i)}\wbf-\bbf^{(i)}}_2^2 + \lambda^2\nrm{\wbf}_0\right\}.
\end{equation}
The linear inequality constraint $\Cbf \wbf \leq \dbf$ encodes the constraints listed in \eqref{fconst}, \eqref{hconst}, and \eqref{dconst} on the forces on $f_\text{a-r}$, $f_\text{align}$, and $f_\text{drag}$, and $\lambda$ is the sparsity threshold. We employ the modified sequential thresholding algorithm from \cite{messenger2020weakpde,messenger2021learning}, with least-squares iterations replaced by solving the associated linearly-constrained quadratic program\footnote{Details on this implementation, in particlar $\Cbf$ and $\dbf$, are given in Appendix \ref{app:csr}.}. 

Since the coefficients $\what^{(i)}$ have no {\it a priori} absolute magnitude, we threshold only on the magnitudes of the given term relative to the response vector $\bbf^{(i)}$, namely, we define the thresholding operator $H_\lambda(\wbf)$ by
\begin{equation}\label{thresh}
H_\lambda(\wbf_j) = \begin{dcases} 0, & \frac{\nrm{\Gbf^{(i)}_j\wbf_j}}{\nrm{\bbf^{(i)}}}\notin[\lambda, \ \lambda^{-1}] \\ \wbf_j, & \text{otherwise}. \end{dcases}
\end{equation}
The constrained sequential thresholding algorithm then consists of alternate between (I) solving \eqref{optprob} with $\lambda=0$ for $\tilde{\wbf}$ with the additional constraint $\supp{\tilde{\wbf}}\subset \supp{\wbf_\ell}$, and (II) setting $\wbf^{(\ell+1)}=H_\lambda(\tilde{\wbf})$. A sweep over $\lambda$ values in the range $10^{-4}$ to $10^0$ is performed according to \cite{messenger2020weak,messenger2021learning} to choose an appropriate threshold $\lambda$.

\subsubsection{Trial Basis Functions}

In the case of the directional force model \eqref{dfm}, we require three bases $\CalF_\text{a-r}$, $\CalF_\text{align}$, and $\CalF_\text{drag}$ for the three proposed forces $f_\text{a-r}$, $f_\text{align}$, and $f_\text{drag}$. We seek a sparse model, and so choose global basis functions, rather than a model composed of a large sum over basis functions that are spatially localized.

For the attractive-repulsive basis $\CalF_\text{a-r}$ we choose products of cosines and scaled and weighted Laguerre polynomials,
\[
\CalF_\text{a-r} = \{\cos(n\theta)p_\ell(\alpha r) e^{-\frac{\alpha}{2}r}\}_{n=0,\ell=0}^{2,17}
\]
for $\ell$th degree Laguerre polynomial $p_\ell$. The scale $\alpha$ is chosen from $r_{\max}$, the maximum observed distance between cells, such that $e^{-\frac{\alpha}{2}r_{\max}} = \ep_\text{mach}\approx e^{-36}$. We set $\alpha=36$ in all cases below since $r_{\max}\approx 2$. 

The pattern of attractive and repulsive regions of the force $f_\text{a-r}$ is not known {\it a priori}, hence the Laguerre basis offers flexibility. The choice of weighted Laguerre polynomials (with weight $\omega(r) = e^{-r/2}$) is guided by the orthogonality relation
\[\int_0^\infty p_m(r)p_n(r)\omega^2(r)dr = \delta_{mn}.\]
We find $\CalF_\text{a-r}$ leads to a well-conditioned matrix $\Gbf$ despite orthogonality not holding with respect to the data distribution.

For the alignment force we choose a basis of shifted cosines and exponential functions,
\[\CalF_\text{align} = \{(1+\cos(n\theta))e^{-2^\ell r}\}_{n=0,\ell=-2}^{2,5}\]
which is informed by the fact that $f_\text{align}$ must be negative. This is easily controlled with $\CalF_\text{align}$ by simply enforcing that the coefficients $\what_\text{align}$ be negative. For the same reason, we choose the drag force from a basis of monomials and cosines, 
\[
\CalF_\text{drag} = \{(1+\cos(n\theta))|v|^\ell\}_{n=0,\ell=0}^{1,4}
\]
as this can also be easily controlled to yield an overall negative $f_\text{drag}$ force by constraining only the basis elements $\what_\text{drag}$. Moreover, monomials capture the physical assumption that resistance to motion should increase with speed.

\subsection{Cluster}\label{step:cluster}

Once the ensemble of models $\CalM$ has been learned, each model is validated on a small neighborhood of cells, and models with poor performance are replaced by a neighboring model with better performance, if one exists. This step can be seen as a form of ``cross-pollination'' that greatly increases the chance of finding accurate models and eliminating inaccurate models. Details on this step can be found in Appendix \ref{app:modelrep}.

Models are then clustered according to the force modes present. Specifically, using the bases above, we can expand each force according to distinct directional modes:
\begin{align*}
f_\text{a-r}(r,\theta) &= f^{(0)}_\text{a-r}(r) + \cos(\theta)f^{(1)}_\text{a-r}(r)+ \cos(2\theta)f^{(2)}_\text{a-r}(r)\\
f_\text{align}(r,\theta) &= f^{(0)}_\text{align}(r) + \cos(\theta)f^{(1)}_\text{align}(r)+ \cos(2\theta)f^{(2)}_\text{align}(r)\\
f_\text{drag}(|v|,\theta) &= f^{(0)}_\text{drag}(|v|) + \cos(\theta)f^{(1)}_\text{drag}(|v|)
\end{align*}
This leads to eight possible force modes, which we order as follows: 
\[\left\{f^{(0)}_\text{a-r},f^{(1)}_\text{a-r},f^{(2)}_\text{a-r},f^{(0)}_\text{align},f^{(1)}_\text{align},f^{(2)}_\text{align},f^{(0)}_\text{drag},f^{(1)}_\text{drag}\right\}.\] 
We associate the sparsity pattern of the force modes with the set of all 8-bit codes, giving a total of $2^8=256$ possible model clusters. Models are partitioned into clusters $\CalC=\{\CalC_1,\dots, \CalC_r\}$ based on their associated codes. For example, species A listed in Table \ref{arttable} below is associated with the code $10111010$ indicating that $f^{(0)}_\text{a-r}$, $f^{(2)}_\text{a-r}$, $f^{(0)}_\text{align}$, $f^{(1)}_\text{align}$, and $f^{(0)}_\text{drag}$ are present in the model.

There are several other options for model replacement and clustering, include clustering based on the sparsity pattern of $\what$, or simply on the presence or absence of each of the three forces $f_\text{a-r}$, $f_\text{align}$, $f_\text{drag}$. The former significantly increases the number of possible clusters, while the latter leads to just 8 possible clusters. Our choice reflects the desire to disentangle directionalities of the governing forces without introducing a strong dependence on the bases used to approximate each force.

\subsection{Aggregate}

Having formed the model clusters $\CalC$, let $\overline{\CalC}$ be the cluster with the most members. We then compute $\overline{\CalM}$ by averaging over the coefficients from models in $\overline{\CalC}$, that is, we compute 
\begin{equation}\label{wbar}
\overline{\wbf} = \frac{1}{|\overline{\CalC}|}\sum_{i\in \overline{\CalC}}\what^{(i)},
\end{equation}
where $|\overline{\CalC}|$ denotes the number of elements in $\overline{\CalC}$. This ensemble average preserves the force modes that identify $\overline{\CalC}$. As explored in \cite{FaselKutzBruntonEtAl2022ProcRSocMathPhysEngSci}, in some cases it may be more appropriate to use the coefficient median, or take a weighted average. We leave these for future work.

\subsection{Validate}\label{sec:validate}

To validate the aggregate model $\overline{\CalM}$, we perform forward simulations over the remaining unclassified cells in a highly  parallelizable way that utilizes the experimental data to efficiently march forward in time.

Let $N'\leq N$ be the number of remaining unclassified cells. For each $i=1,\dots,N'$, we simulate a new trajectory $\{(\overline{x}_i(t_k),\overline{v}_i(t_k))\}_{k=1}^L$ using the averaged model $\overline{\CalM}$ with the experimental initial conditions $(\overline{x}_i(0),\overline{v}_i(0))=(\xbf_i(0),\vbf_i(0))$.  We march forward in time according to the forward Euler update:
\begin{align}
\overline{v}_i(t_{k+1}) &= \overline{v}_i(t_k) + \Delta t \overline{\CalM}(\overline{x}_i(t_k),\overline{v}_i(t_k),\Xbf^{\prime i}(t_k),\Vbf^{\prime i}(t_k))\\
\overline{x}_i(t_{k+1}) &= \overline{x}_i(t_k) + \Delta t\overline{v}_i(t_k)
\end{align}
where $(\Xbf^{\prime i}(t_k),\Vbf^{\prime i}(t_k))$ indicates $(\Xbf(t_k),\Vbf(t_k))$ with the $i$-th cell removed\footnote{By some abuse of notation, $\overline{\CalM}(x,v,X,V)$ is used to denote the instantaneous force on a particle $(x,v)$ from neighboring cells $(X,V)$ under model $\overline{\CalM}$}. Since the time resolution of the data $\Delta t$ is assumed to be coarse, we perform the simulation on a finer grid with timestep $\Delta t'=2^{-5}\Delta t$, and use piecewise cubic hermite interpolation to generate positions and velocities  of neighbors $(\Xbf^{\prime i},\Vbf^{\prime i})$ at intermediate times. We stress that we do not update the neighbor cells using the model $\overline{\CalM}$, which would be much more costly, we merely use neighbor positions and velocities from the data to compute interactions that govern the motion of cell $i$. The resulting trajectories $\{(\overline{x}_i,\overline{v}_i)\}_{i=1}^{N'}$ can then be trivially computed in parallel\footnote{See Table \ref{table:time} for simulation walltimes.}.

We then define the validation error for cells $i=1,\dots,N'$ as the relative velocity difference 
\begin{equation}\label{VEformula}
\Delta V_i:=\sqrt{\frac{\sum_{k=1}^{L'} \nrm{\overline{v}_i(t_k)-\vbf_i(t_k)}_2^2}{\sum_{k=1}^{L'} \nrm{\vbf_i(t_k)}_2^2}},
\end{equation}
where $L'\leq L$ is a subset of the time series over which the simulation is expected to remain accurate\footnote{In particular, for chaotic systems the trajectories cannot be expected to remain close for large times, however the correct model will be initially accurate.}. In this work we choose $L' = 0.25L$. With $L=200$ timesteps in the examples below, this results in a comparison with the data over the first 50 timesteps at the original scale $\Delta t$, or equivalently 1600 timesteps on the finer scale $\Delta t'$.

\subsection{Classify}

\begin{figure*}
\centering
\begin{tabular}{ccc}
	\hspace{-0.25cm} \includegraphics[trim={35 0 40 15},clip,width=0.31\textwidth]{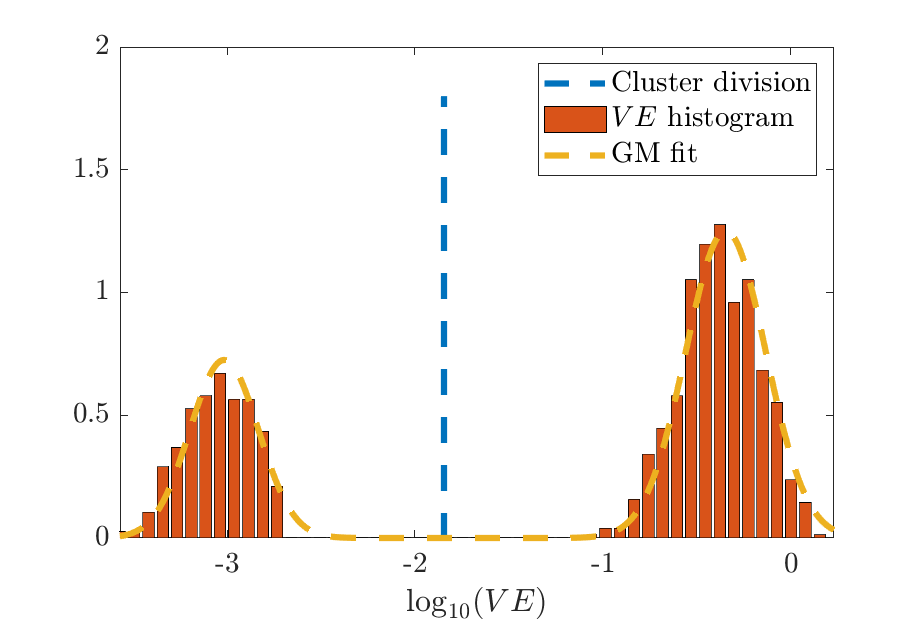} & 		\hspace{-0.1cm} \includegraphics[trim={35 0 40 15},clip,width=0.31\textwidth]{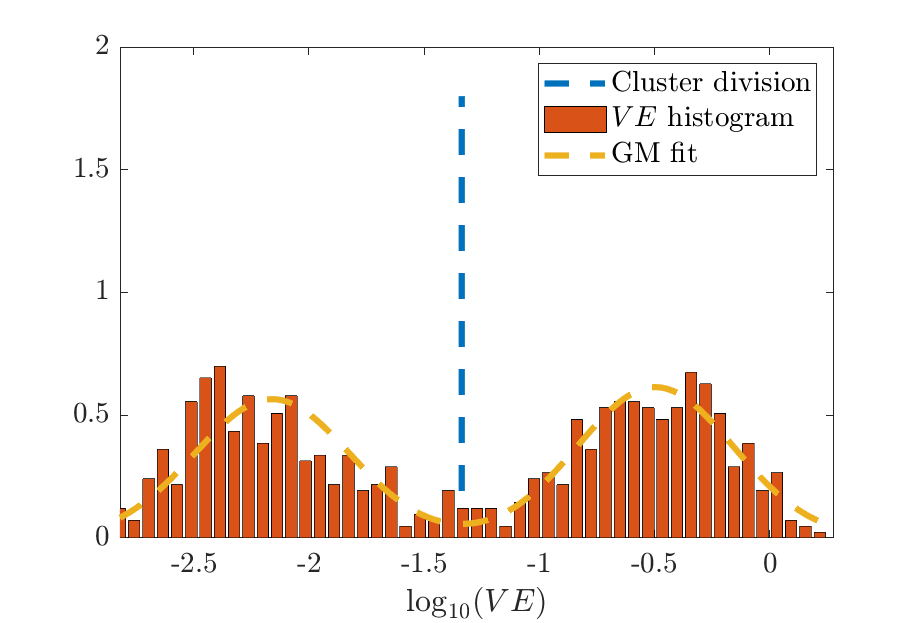} &
\hspace{-0.1cm}    \includegraphics[trim={35 0 40 15},clip,width=0.31\textwidth]{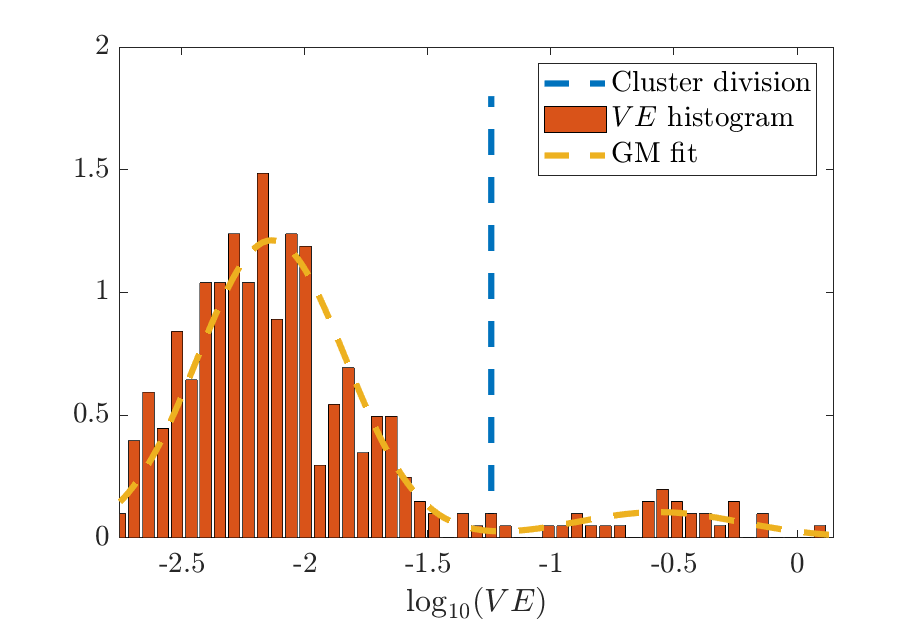} 
\end{tabular}
\caption{Gaussian mixture models for classifying the three-species experiment $\Xbf_{A,B,C}$(long) (see Table \ref{tableXABC} row 3 for details). We see an initial complete separation of species $C$ (left), followed by a mixed cluster containing $94\%$ of the species B cells and $1.8\%$ of the species $A$ cells (middle). The next iteration classifies a majority species A cluster (right). Clusters 4 and 5 are effectively outliers and contain the remaining 31/1000 cells.}
\label{gmmXABC}
\end{figure*}

Let $VE$ be the set of validation errors, $VE = \{\Delta V_1, \dots,\Delta V_{N'}\}$. An empirical observation used in this work is that when $\overline{\CalM}$ approximates well an underlying model for a true species, the log-transformed validation errors $\log_{10}(VE)$ are fit well by a Gaussian mixture model (GMM) with two mixtures (see Figures \ref{gmmXACXBC},\ref{gmmXAB},\ref{gmmXABC}). We thus use a 2-mixture GMM to classify the remaining cells. Cells are granted membership into the mixture that yields the highest posterior probability of generating its log-validation error, and the class with lowest mean error is labeled as a species. This can thought of as a sequential binary classification scheme.

For example, in each plot of Figure \ref{gmmXACXBC}, a representative GMM resulting from a two-species population, the left-most mixture corresponds to low validation errors under the model $\overline{\CalM}$ and is classified as a species $\overline{\CalS}$ (in this case, species C in Table \ref{arttable} is identified). The cells in $\overline{\CalS}$ are subsequently removed from the population, and cells in the right-most mixture are returned to the cluster phase (b). \\

\subsection{Stopping conditions}

Steps (b)-(e) are repeated until one the following conditions is met.
\begin{enumerate}[label=(\arabic*)]
\item Less than 3 cells remain: $N'\leq 2$
\item More than $99\%$ of remaining cells have less than $5\%$ validation error: $\#\{VE<0.05\}\geq 0.99N'$
 \item More than 10 species have been identified: $S>10$
\end{enumerate}

The first case is an obvious criterion to prevent infinite looping over outlier cells for which there is not enough information to learn an adequate model. The second condition skips the GMM fitting process when all cells have sufficiently low error\footnote{We choose this cutoff at $5\%$ but note that this parameter may need tuning.}, and directly assigns all unlabeled cells to a new species. This is necessary to account for the cases of {\it high-accuracy} recovery, where it is observed that the $VE$ is no longer approximately log-normal. For example, this is observed in the rightmost plot of Figure \ref{gmmXAXBXC}. Finally, for $N$ very large, it may be necessary to restrict the total number of species, which is encapsulated in the third condition, although we did not observe the number of iterations exceeding 5 in any trials with $N\leq 1000$ and $S\leq 3$ true species.


   
\section{Results: Artificial Cells}\label{sec:results}

\begin{figure*}
\begin{center}
\begin{tabular}{cc}
		\includegraphics[trim={25 80 25 25},clip,width=0.45\textwidth]{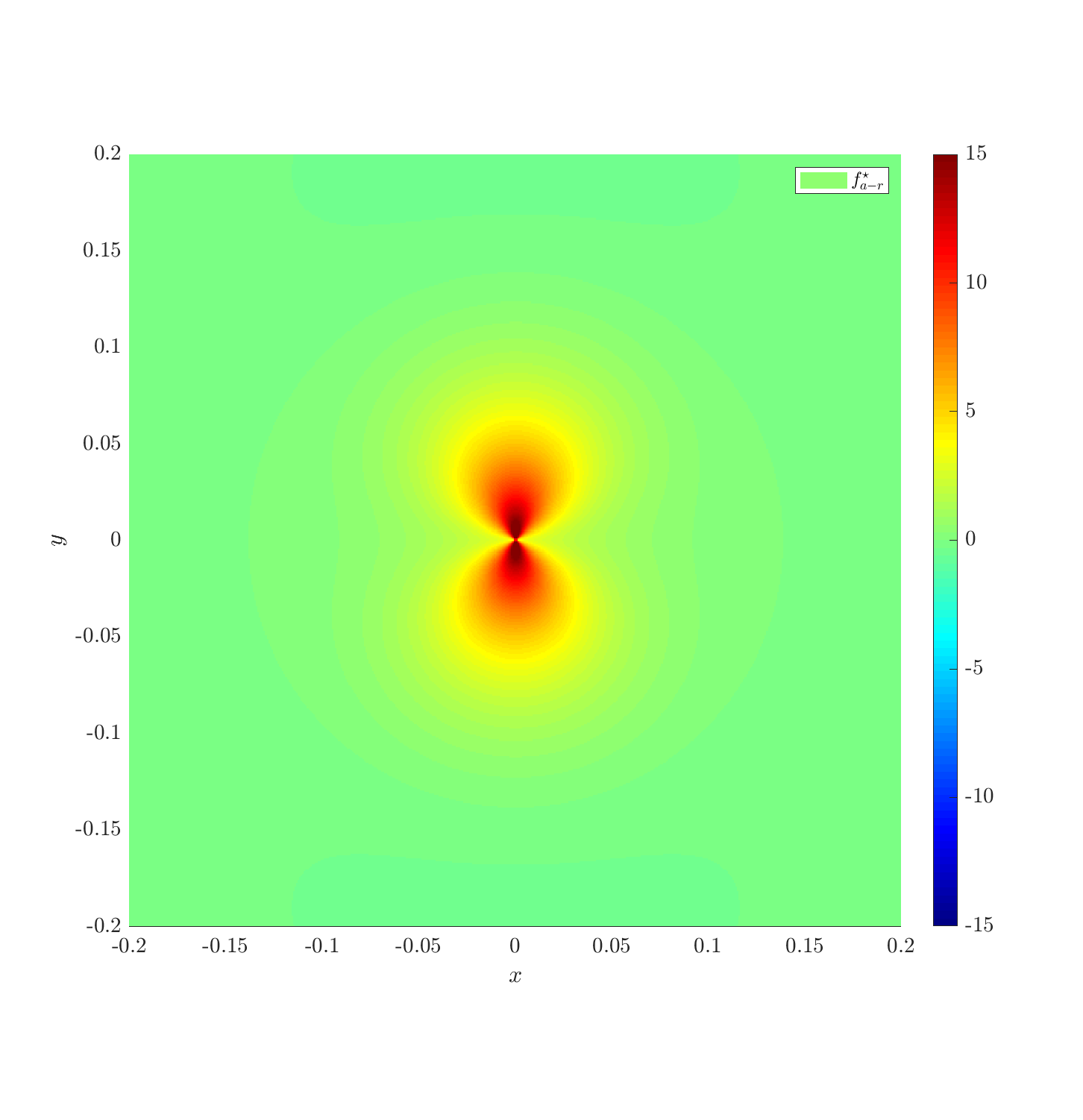} & 
		\includegraphics[trim={25 80 25 25},clip,width=0.45\textwidth]{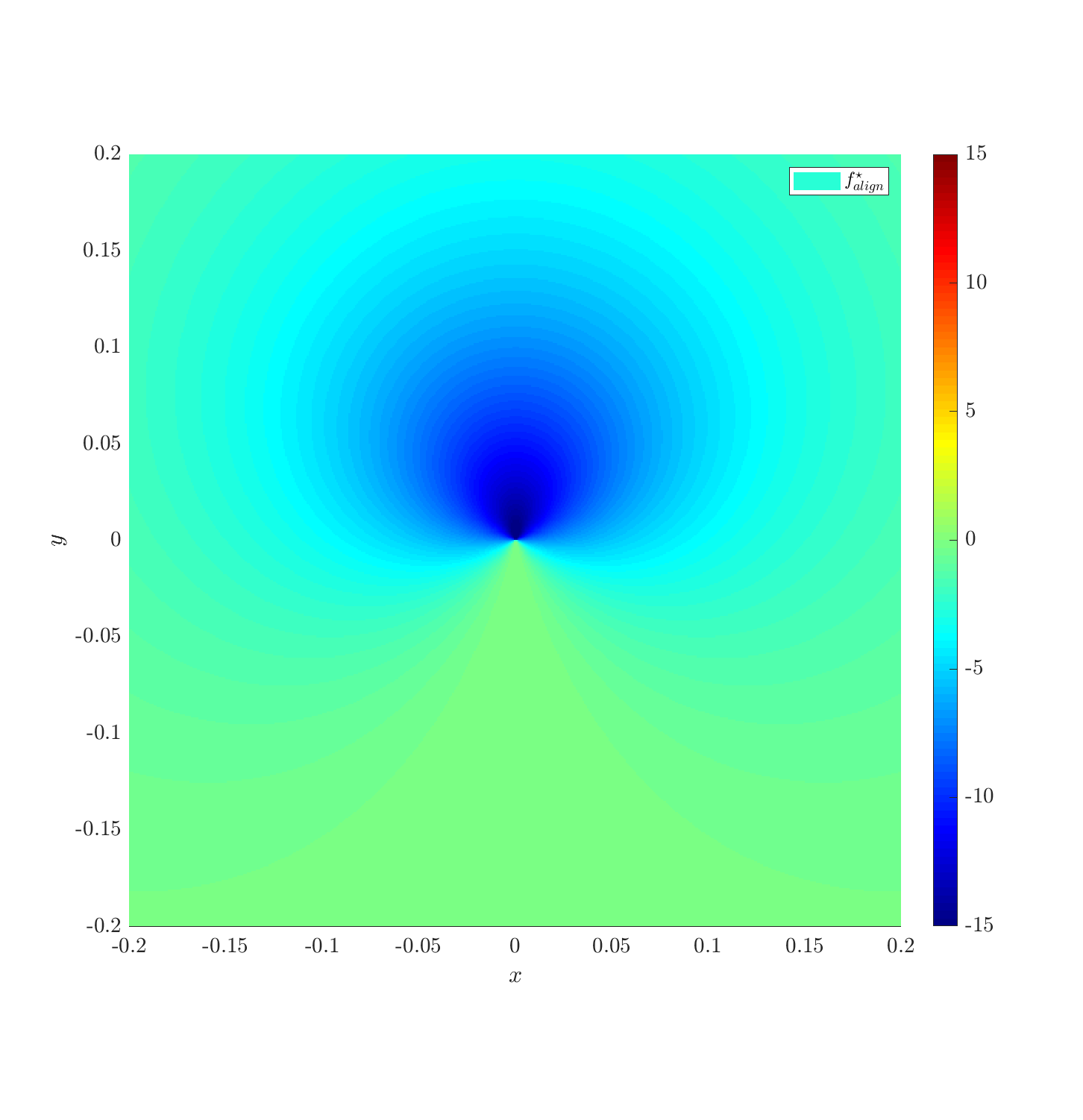}
\end{tabular}
\end{center}
\caption{Forces used to generate artificial data, motivated by experiment. Left: quadrupolar attractive-repulsive force $f^\star_\text{a-r}$. Right: dipolar alignment force $f^\star_\text{align}$. Note that all artificial species are prescribed the same drag force $f^\star_\text{drag}(s, \theta) = -5s$, linear in cell speed $s$.}
\label{artforce}
\end{figure*}

\begin{table*}

\begin{subequations}
\[
\boxed{\begin{array}{rl}
f_{\text{a-r}}^{\star}(r,\theta) & :=(15+10\cos(2\theta))\left(e^{-20r}-0.25e^{-10r}\right)\\
f_{\text{align}}^{\star}(r,\theta) & :=-(8+8\cos(\theta))e^{-8r}\\
f_{\text{drag}}^{\star}(s,\theta) & :=-5s
\end{array}}
\]
\end{subequations}
\begin{center}
    \begin{tabular}{|c|c|c|c|}
    \hline
    & $f^\star_\text{a-r}$ &  $f^\star_\text{align}$ &  $f^\star_\text{drag}$ \\ 
    \hline
    \cellcolor[rgb]{0.7608,0.2784,1.0000}{Species A} & $\checkmark$ & $\checkmark$ & $\checkmark$ \\ 
    \hline
    \cellcolor[rgb]{0,1,1}{Species B} & $\checkmark$ &  & $\checkmark$ \\ 
    \hline
    \cellcolor[rgb]{1,0,0}{Species C} & & $\checkmark$ & $\checkmark$ \\ 
    \hline
    \end{tabular}
\end{center}

\caption{Species delineation by active force modes.}
\label{arttable}
\end{table*}

We examine artificial cell cultures with combinations of 1-3 distinct cell types, denoted by species A, species B, and species C. Each species has a unique combination of the forces depicted in Figure \ref{artforce}, with the combinations and force definitions specified in Table \ref{arttable}. The forces include a {\it quadrupolar} attractive-repulsive force $f^\star_\text{a-r}$, a {\it dipolar} alignment force $f^\star_\text{align}$, and a {\it monopolar} drag force $f_\text{drag}^\star$ which is linear in its speed argument. As we will see below, species A and species B share the force $f_\text{a-r}$ and hence result in similar dynamics, which presents a challenge to identification. It turns out that using a longer time series results in correct classification.

 We let $\Xbf_P$ denote a simulation with individuals from species $P$, for example, $\Xbf_A$ is a simulation with only individuals from species A and $\Xbf_{A,B}$ is a simulation with a mixed population of species A and species B. Each simulation has 1000 individuals and an even number of members in each species (up to rounding). More details on the simulations can be found in the appendix.

We are particularly interested in three traits of our learning algorithm: (1) was the classification successful? (2) are the learned forces close to the true forces? (3) are simulated trajectories using the learned model close to the original trajectories? To assess (1) we report the {\it classification success} CS$(i)$ for $i\in \{A,B,C\}$ as the fraction of individual from species $i$ that ended up in the cluster in question, where clusters are listed as subrows (rows not separated by horizontal lines) within each row in Tables \ref{tableXAXBXC}, \ref{tableXACXBCXAB}, and \ref{tableXABC}, in the order they were identified in step (e) of the algorithm. For example, in row 2 of Table \ref{tableXACXBCXAB}, two clusters are identified from the two-species data $\Xbf_{A,C}$, with the first cluster containing $100\%$ of the species C cells, indicated by CS$(C)=1.000$, and the second cluster containing $100\%$ of the species A cells, indicated by CS$(A)=1.000$, with no outliers.

To assess the accuracy of learned forces with respect to the ground truth forces, for each of the three forces we report the relative $L^2$ error over $r\in [0,2]$ (since $r_{\max}\approx2$ for all examples here) and $\theta\in\{0, \pi/4, \pi/2, 3\pi/4, \pi\}$, denoting this by $\Delta f_\text{a-r}$, $\Delta f_\text{align}$, and $\Delta f_\text{drag}$.

Lastly, we assess the difference in learned and true trajectories using the average validation error $\Delta V = \frac{1}{|\overline{\CalS}|}\sum_{i\in \overline{\CalS}} \Delta V_i$, where $\Delta V_i$ is computed from model $\overline{\CalM}$ associated with identified species $\overline{\CalS}$ using \eqref{VEformula}.

\subsection{Homogeneous Populations}

\begin{table*}
\begin{center}
\begin{tabular}{|c|c|c|c|c|c|r|c|}
\hline
Experiment & $\Delta f_\text{a-r}$ & $\Delta f_\text{align}$ &  $\Delta f_\text{drag}$ & CS($A$)  & CS($B$)  & CS($C$) & $\Delta V$ \\
\hline
$\Xbf_A$ & $0.0211$ & $0.0384$ & $0.0382$ & \cellcolor[rgb]{0.7608,0.2784    1.0000}{1.000} & --- & --- & 0.0100 \\
\hline
$\Xbf_B$ & $0.0098$ & --- & $0.0125$ & --- & \cellcolor[rgb]{0,1,1}{0.997} & --- & 0.0076 \\
\hline
$\Xbf_C$ & --- & $0.0011$  & $0.0016$ & --- & --- & \cellcolor[rgb]{1,0,0}{1.000} & 0.0016 \\ \hline
\end{tabular}
\end{center}
\caption{Performance of model learning and classification algorithm of homogeneous populations.}
\label{tableXAXBXC}
\end{table*}

\begin{figure*}
\begin{tabular}{ccc}
	\hspace{-0.25cm} \includegraphics[trim={40 5 40 15},clip,width=0.35\textwidth]{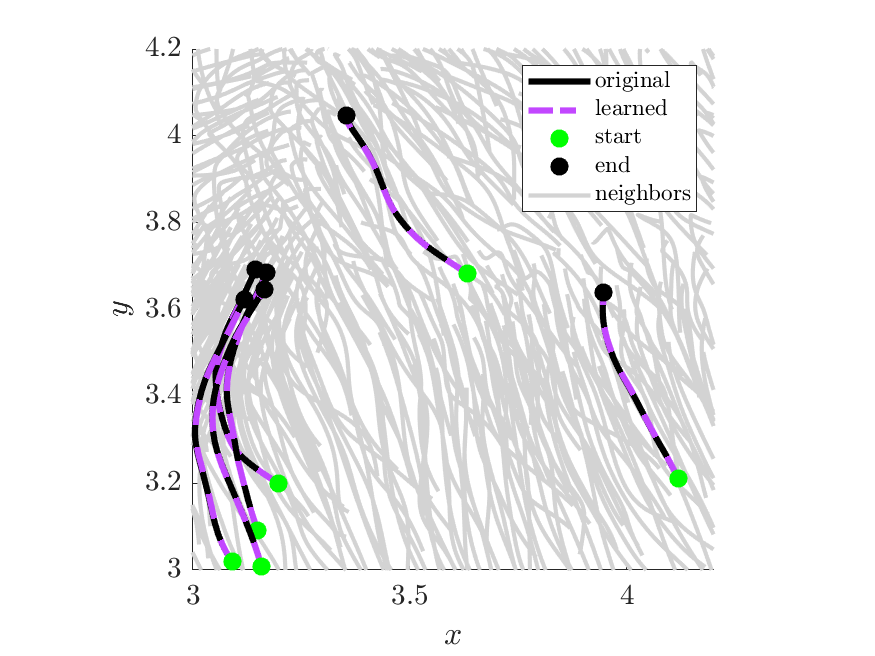} & 		\hspace{-0.9cm} \includegraphics[trim={40 5 40 15},clip,width=0.35\textwidth]{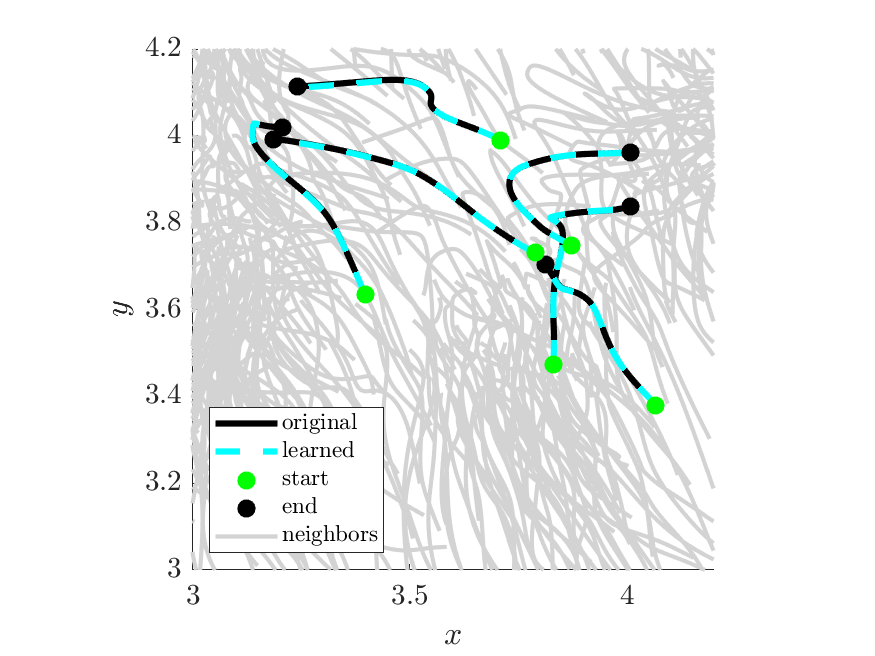} &
\hspace{-0.9cm}    \includegraphics[trim={40 5 40 15},clip,width=0.35\textwidth]{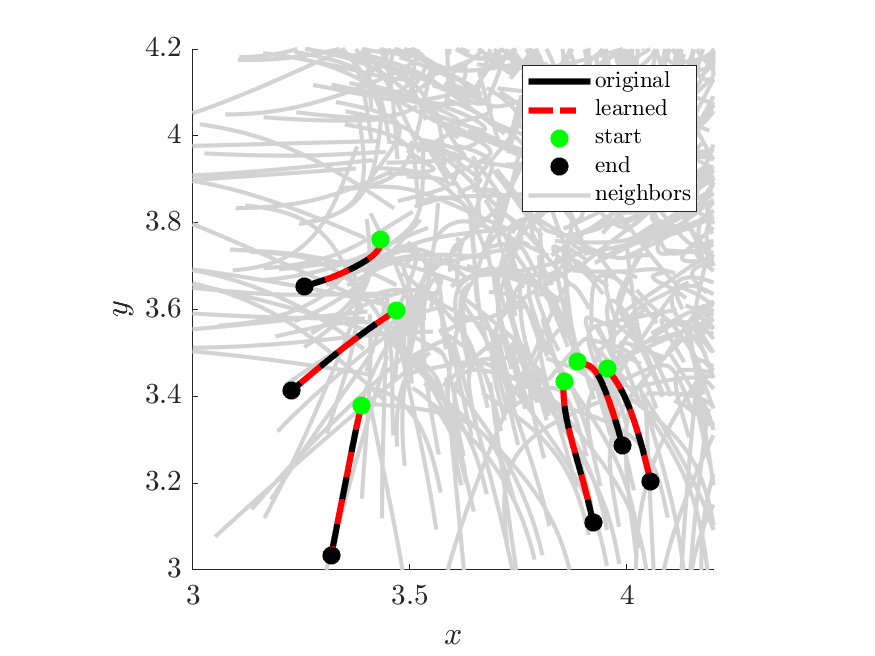} 
\end{tabular}
\caption{Examples of learned and original trajectories from homogeneous populations. Left to right: $\Xbf_A$, $\Xbf_B$, $\Xbf_C$.}
\label{trajXAXBXC}
\end{figure*}

\begin{figure*}
\begin{tabular}{ccc}
		\includegraphics[trim={20 0 20 0},clip,width=0.31\textwidth]{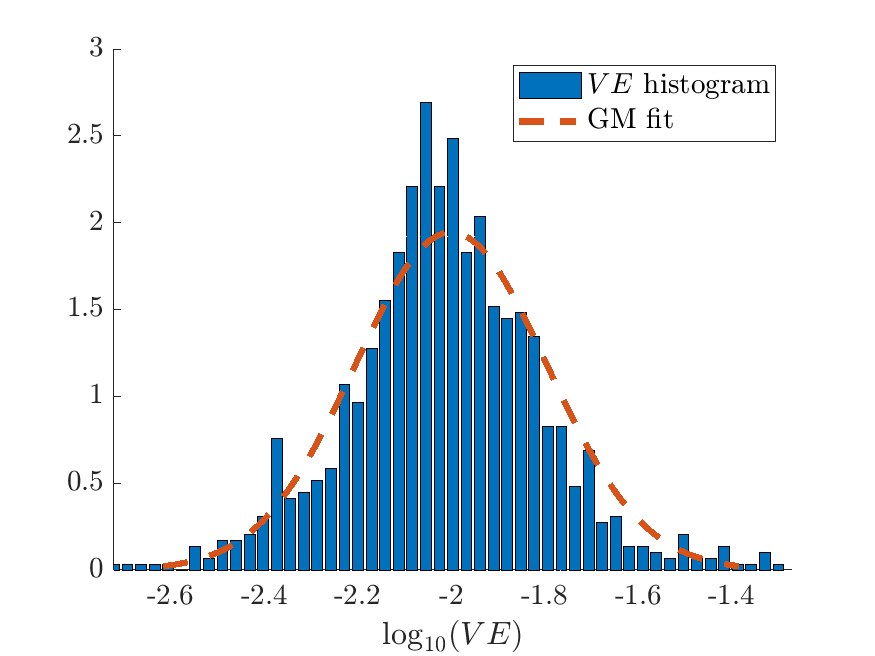} & 
		\includegraphics[trim={20 0 20 0},clip,width=0.31\textwidth]{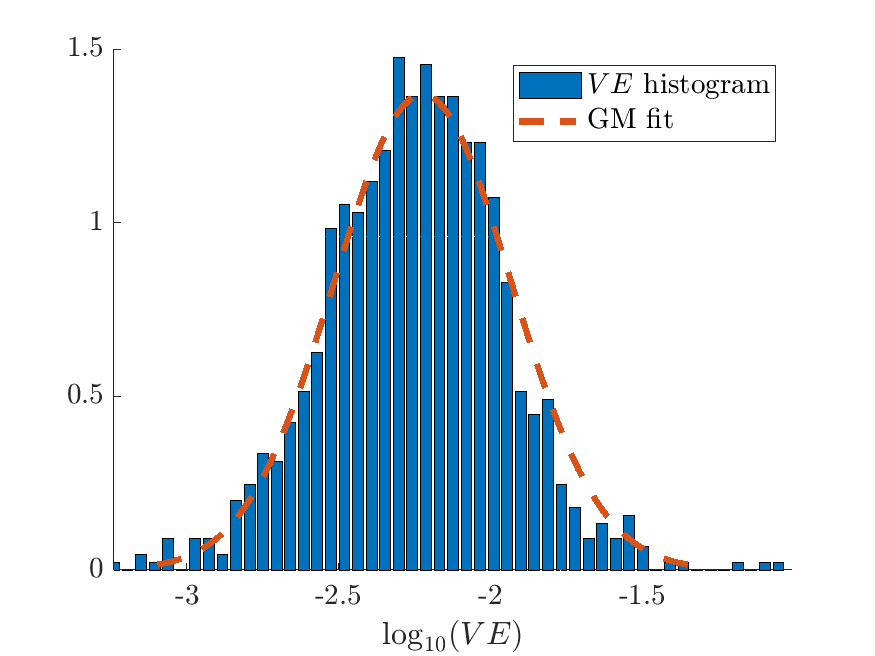} &
		\includegraphics[trim={20 0 20 0},clip,width=0.31\textwidth]{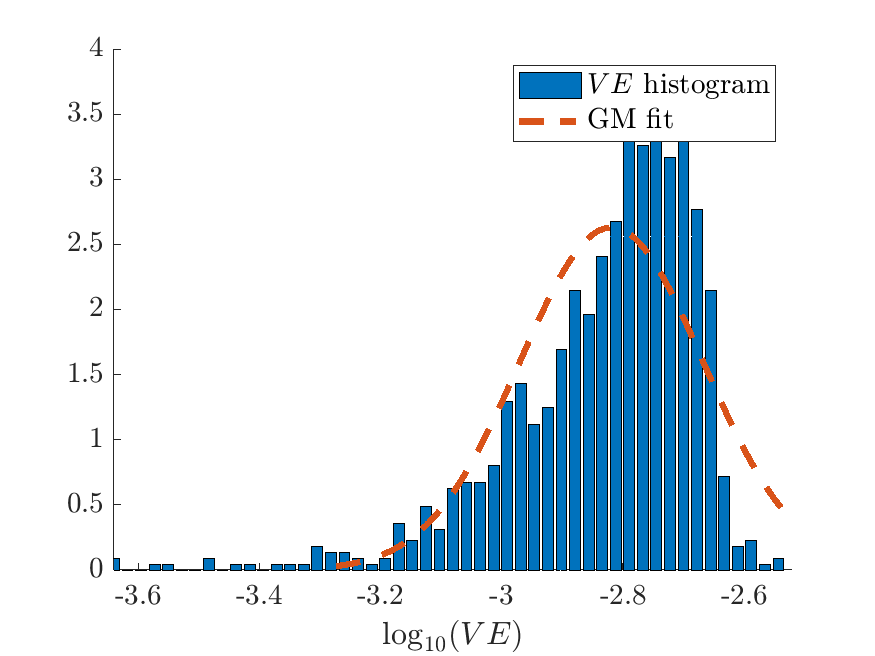}
\end{tabular}
\caption{Distribution of log-validation errors for homogeneous cell experiments $\Xbf_A$, $\Xbf_B$, $\Xbf_C$. Distributions for $\Xbf_A$ and $\Xbf_B$ are fit well by a single Gaussian, indicating a single species is present. The distribution for $\Xbf_C$ has a non-Gaussian tail, although all errors are below $1\%$, indicating that the candidate model fits the population up to the specified error tolerance.}
\label{gmmXAXBXC}
\end{figure*}

As an initial benchmark we detect single-species populations from homogeneous data. While simpler than the heterogeneous case, this is a nontrivial task due to the variability of single-cell trajectories and local environments within the population. Our method successfully identifies the models for species $A$, $B$, and $C$ from homogeneous simulations, achieving less than $1\%$ mean validation errors in each case, and less than $4\%$ relative force errors $\Delta f$ (Table \ref{tableXAXBXC}). In simulation $\Xbf_B$, three cells are identified as outliers (appearing in the right tail of Figure \ref{gmmXAXBXC} (middle)), and all other cells in $\Xbf_A$, $\Xbf_B$, and $\Xbf_C$ are correctly classified. A comparison between original and learned trajectories is depicted in Figure \ref{trajXAXBXC}, with learned trajectories overlapping original trajectories in each case.

\subsection{Two-Species Populations}

\begin{table*}
\begin{center}
\begin{tabular}{|c|c|c|c|c|c|r|c|}
\hline Experiment & $\Delta f_\text{a-r}$ & $\Delta f_\text{align}$ &  $\Delta f_\text{drag}$ & CS($A$)  & CS($B$)  & CS($C$) & $\Delta V$ \\
\hline $\Xbf_{A,C}$ &
\begin{tabular}{c}
--- \\ 
$0.0226$ \end{tabular} & 
\begin{tabular}{c}
$0.0011$ \\ 
$0.0941$\end{tabular} & 
\begin{tabular}{c}
$0.0002$ \\
$0.0472$ \end{tabular}  &
\begin{tabular}{c}
0\\ 
\cellcolor[rgb]{0.7608,0.2784,1.0000}{1.000} \end{tabular} & 
\begin{tabular}{c}
--- \\ 
---\end{tabular} & 
\begin{tabular}{c}
\cellcolor[rgb]{1,0,0}{1.000}\\ 
0 \end{tabular} & 
\begin{tabular}{c}
0.0005 \\ 
0.0200 \end{tabular}\\
\hline
$\Xbf_{B,C}$ & 
\begin{tabular}{c}
--- \\ 
$0.0328$ \end{tabular} & 
\begin{tabular}{c}
$0.0077$ \\ 
---\end{tabular} & 
\begin{tabular}{c}
$0.0051$ \\
$0.0461$ \end{tabular}   &
\begin{tabular}{c}
--- \\ 
--- \end{tabular} & 
\begin{tabular}{c}
0 \\ 
\cellcolor[rgb]{0,1,1}{1.000}\end{tabular} & 
\begin{tabular}{c}
\cellcolor[rgb]{1,0,0}{1.000}\\ 
0 \end{tabular} & 
\begin{tabular}{c}
0.0023 \\ 
0.0339 \end{tabular}\\ 
\hline
$\Xbf_{A,B}$ & 
\begin{tabular}{c}
$0.0341$ \\ 
$0.0075$ \\ 
$0.4780$ \end{tabular} & 
\begin{tabular}{c}
--- \\ 
--- \\ 
$0.3660$ \end{tabular} & 
\begin{tabular}{c}
$0.0133$ \\
$0.0538$ \\ 
--- 
\end{tabular} &
\begin{tabular}{c}
0.102 \\ 
0.084 \\ 
0.814 \end{tabular} & 
\begin{tabular}{c}
0.628 \\ 
0.340 \\ 
0.012 \end{tabular} & 
\begin{tabular}{c}
--- \\ 
--- \\ 
--- \end{tabular} & 
\begin{tabular}{c}
0.1201 \\ 
0.0362 \\ 
0.3945 \end{tabular}\\
\hline
$\Xbf_{A,B}$(long) & 
\begin{tabular}{c}
$0.0018$ \\ 
$0.0034$ \\ 
$0.0071$ \\
$0.0034$
\end{tabular} & 
\begin{tabular}{c}
--- \\ 
0.0023 \\ 
--- \\
0.0051 \end{tabular} & 
\begin{tabular}{c}
0.0042 \\
0.0067 \\ 
0.0044\\
0.0144
\end{tabular} &
\begin{tabular}{c}
0.002 \\ 
\cellcolor[rgb]{0.7608,0.2784,1.0000}{0.994} \\
0 \\
0.004 \end{tabular} & 
\begin{tabular}{c}
\cellcolor[rgb]{0,1,1}{0.978} \\ 
0 \\ 
0.023 \\
0 \end{tabular} & 
\begin{tabular}{c}
--- \\ 
--- \\ 
--- \\
---\end{tabular} & 
\begin{tabular}{c}
0.0045 \\ 
0.0070 \\ 
0.0568 \\
0.0199 \end{tabular}\\
\hline
\end{tabular}
\end{center}
\caption{Performance of model learning and classification algorithm for two-species populations. $\Xbf_{A,B}$(long) is simply the continuation of $\Xbf_{A,B}$ to twice the time horizon, and significantly improves classification over $\Xbf_{A,B}$.}
\label{tableXACXBCXAB}
\end{table*}

\begin{figure*}
\begin{center}
\begin{tabular}{cc}
		\includegraphics[trim={0 0 0 0},clip,width=0.45\textwidth]{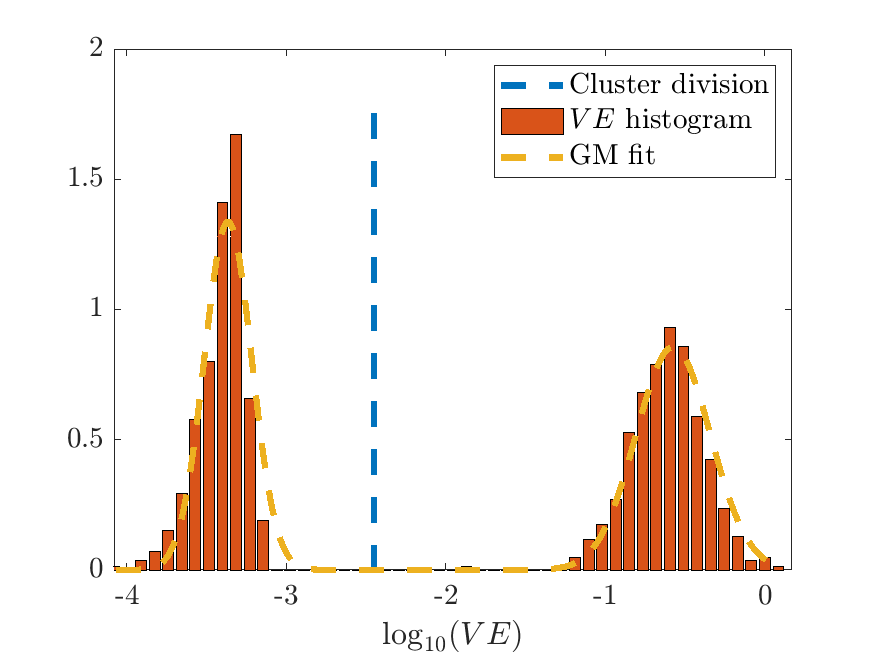} &
		\includegraphics[trim={0 0 0 0},clip,width=0.45\textwidth]{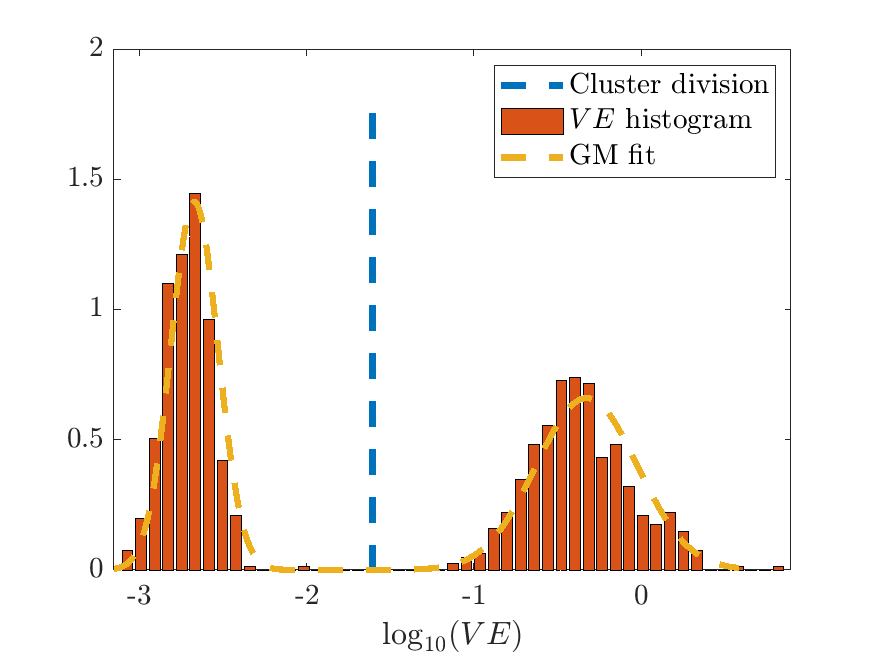} 
\end{tabular}
\end{center}
\caption{Distribution of log-validation errors for heterogeneous cell experiments $\Xbf_{A,C}$ and $\Xbf_{B,C}$. In each case, species $C$ is identified in the first iteration, and a clear separation between the two species allows for accurate clustering using Gaussian mixture models.}
\label{gmmXACXBC}
\end{figure*}

\begin{figure*}
\begin{tabular}{ccc}
	\hspace{-0.25cm} \includegraphics[trim={40 5 40 15},clip,width=0.34\textwidth]{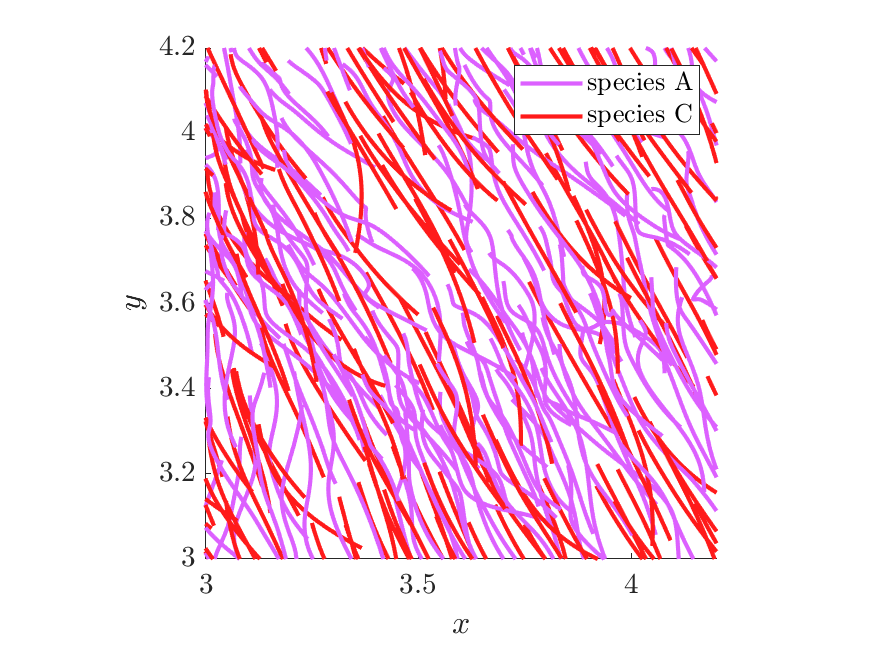} & 		\hspace{-0.9cm} \includegraphics[trim={40 5 40 15},clip,width=0.34\textwidth]{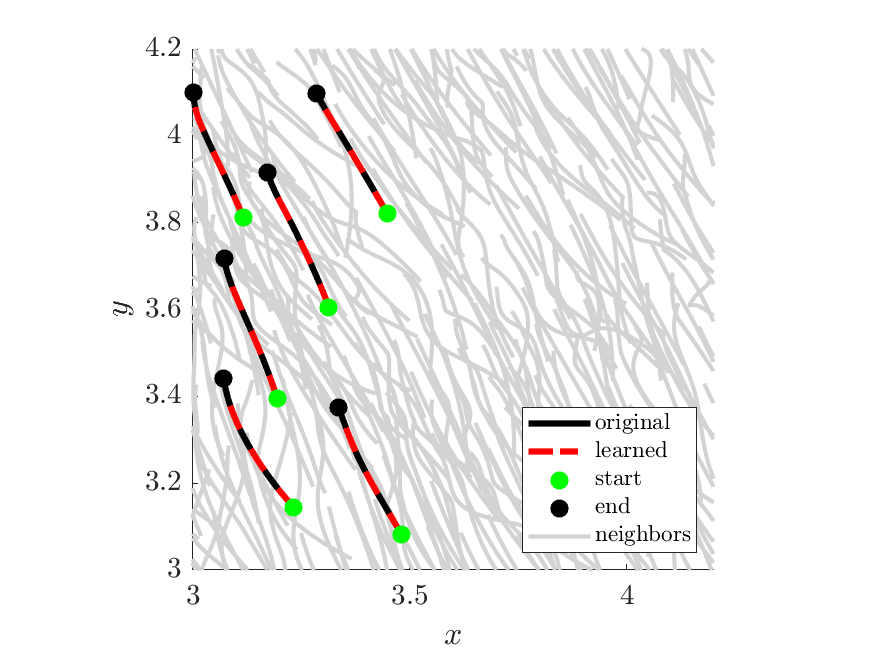} &
\hspace{-0.9cm}    \includegraphics[trim={40 5 40 15},clip,width=0.34\textwidth]{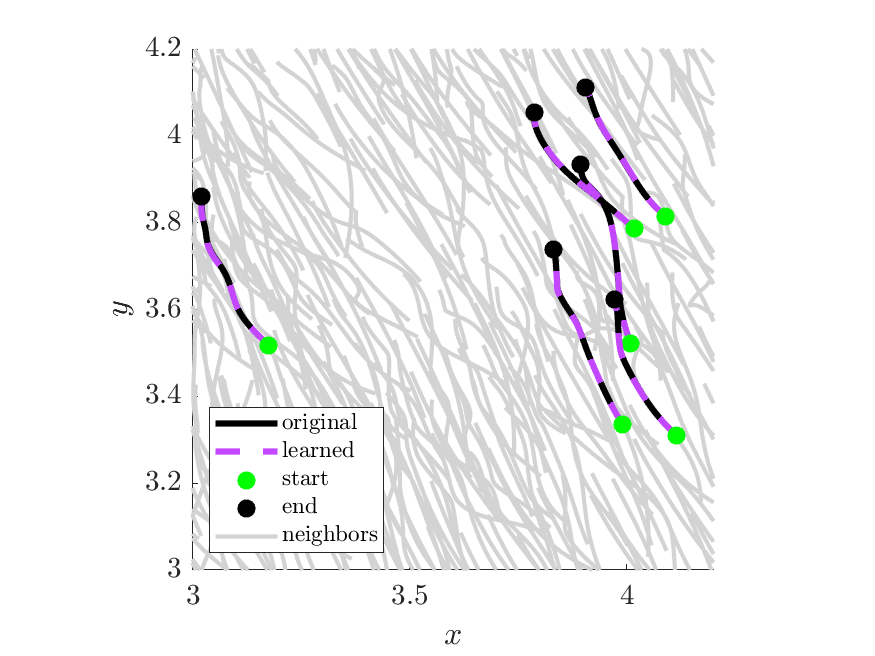} 
\end{tabular}
\caption{Example trajectories from experiment $\Xbf_{A,C}$. Cells with true color labels are depicted on the left, but are passed into the algorithm unlabeled. The algorithm then classifies the population into different species and returns accurate models for each species. Classified cells from species $C$ (middle) and species $A$ (right) are highlighted showing excellent agreement between data and simulation.}
\label{trajXAC}
\end{figure*}

\begin{figure*}
\begin{tabular}{ccc}
	\hspace{-0.25cm} \includegraphics[trim={40 5 40 15},clip,width=0.34\textwidth]{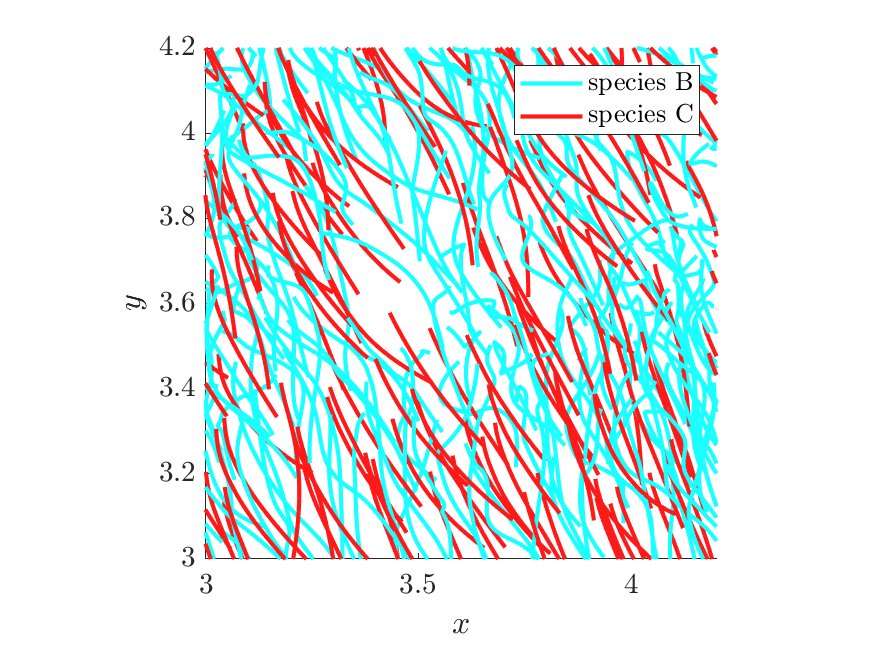} & 		\hspace{-0.9cm} \includegraphics[trim={40 5 40 15},clip,width=0.34\textwidth]{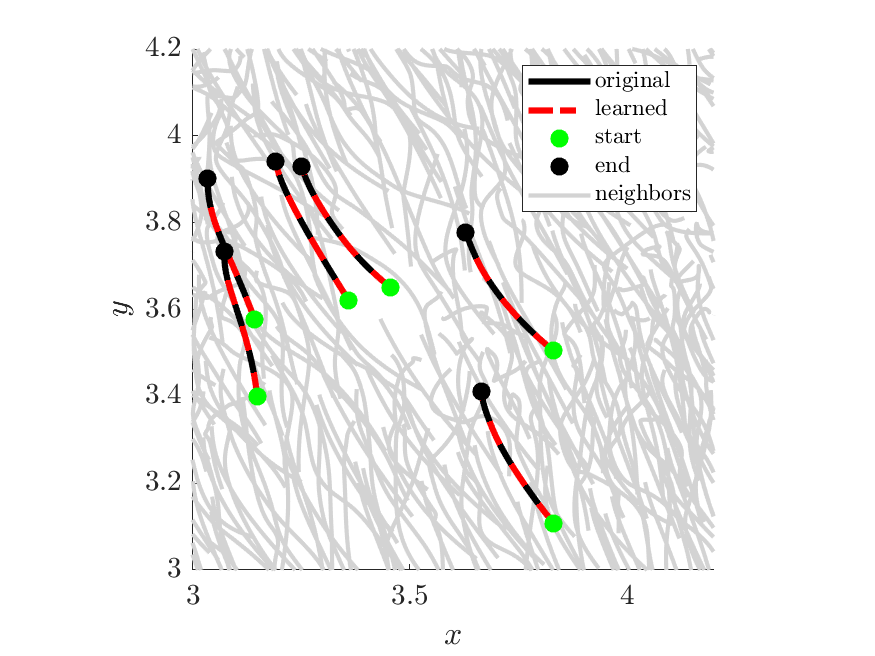} &
\hspace{-0.9cm}    \includegraphics[trim={40 5 40 15},clip,width=0.34\textwidth]{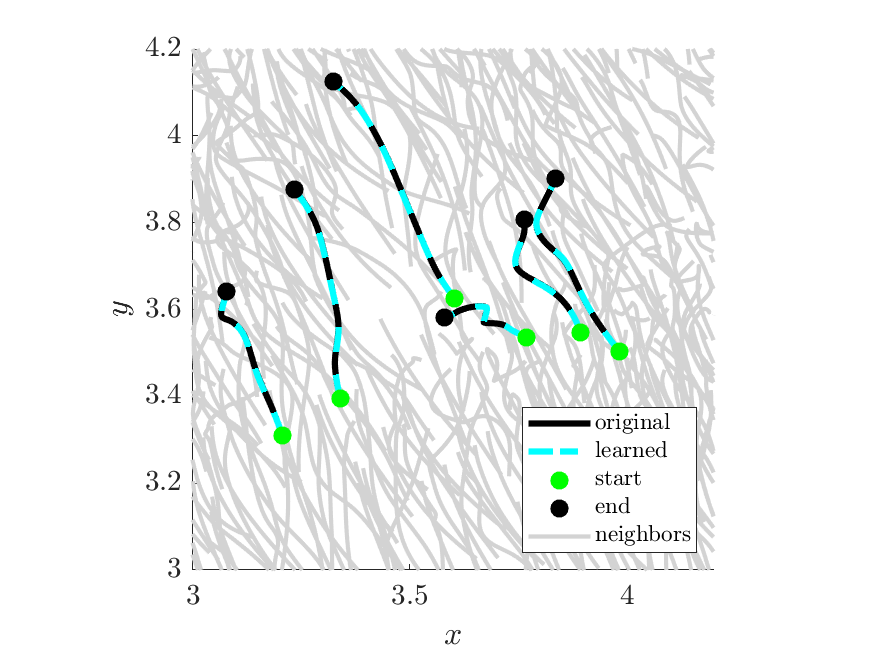} 
\end{tabular}
\caption{Example trajectories from experiment $\Xbf_{B,C}$. As in Figure \ref{trajXAC},  cells with true color labels are depicted on the left. Classified cells from species $C$ (middle) and species $B$ (right) are highlighted showing excellent agreement original data and output of the learned models.}
\label{trajXBC}
\end{figure*}

Next we examine the ability of the learning algorithm to detect two-species populations along with accurate aggregate models. Figure \ref{gmmXACXBC} displays two representative Gaussian mixture fits to the log-validation errors for $\Xbf_{A,C}$ (left) and $\Xbf_{B,C}$ (right). In both cases, the log-errors are well-approximated by Gaussian mixtures with wide separations between mixtures. This allows for complete classification in both cases, as indicated by CS$(A)$, CS$(B)$, and CS$(C)$ in rows 2 and 3 of Table \ref{tableXACXBCXAB}. Force differences $\Delta f$ are less than $5\%$ in each case, with trajectory validation errors less than $3.5\%$. In particular, species C achieves less that $0.3\%$ validation error, which is due to the true force $f_\text{align}^\star$ existing in the learning library, whereas $f_\text{a-r}^\star$ is approximated using a truncated series expansion, resulting in larger errors. See Figures \ref{trajXAC} and \ref{trajXBC} for comparison between original and learned trajectories.

For experiment $\Xbf_{A,B}$, initially the method is incapable of correctly classifying cells into species A and species B. Three clusters are identified with sub-optimal models (Table \ref{tableXACXBCXAB} row 4). Accurate classification is achieved by running the algorithm with a longer experiment $\Xbf_{A,B}$(long) (Table \ref{tableXACXBCXAB} row 5) which is the continuation of $\Xbf_{A,B}$ for twice the total time points, at the same temporal resolution. An initial cluster is identified containing $97.8\%$ of species $B$ along with $0.2\%$ (a single cell) of the existing species $A$ cells, followed by a second cluster with $99.4\%$ of the species $A$ cells and no cells from species $B$. The last two clusters correctly partition the remaining cells (12 in total), again finding accurate models, allowing for recombination with the first two cluster during post-processing. 

\begin{figure*}
\begin{center}
\begin{tabular}{cc}
\includegraphics[trim={0 0 0 0},clip,width=0.33\textwidth]{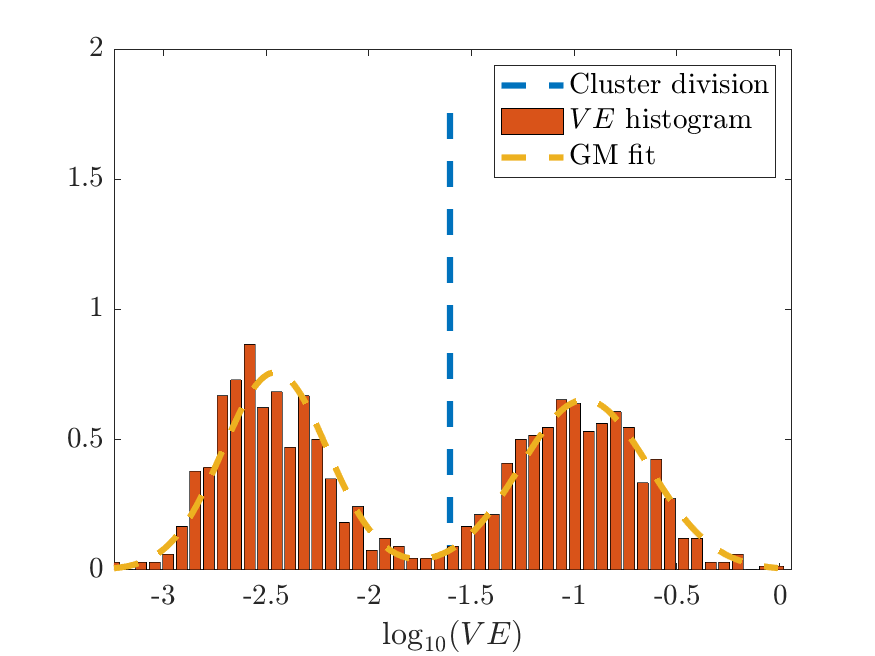} &
\includegraphics[trim={0 0 0 0},clip,width=0.33\textwidth]{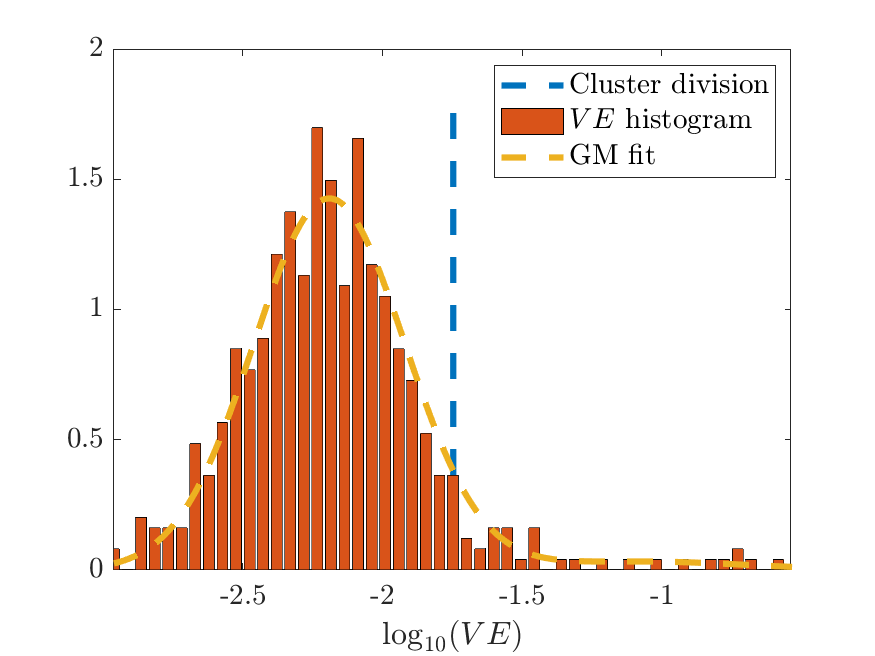} \\
\end{tabular}
\end{center}
\caption{Log-validation errors for two-species population $\Xbf_{A,B}$(long). Strong similarities between the two species present an initial challenge to identification, which is overcome by taking a longer time series. The initial Gaussian mixture model (left) identifies a majority species B cluster. In the second iteration (right), a cluster with all Species A cells is identified, and a small group of cells remains which is then partitioned correctly (see row 5, columns CS(A) and CS(B) of Table \ref{tableXACXBCXAB}).}
\label{gmmXAB}
\end{figure*}

\begin{figure*}
\begin{tabular}{ccc}
	\hspace{-0.25cm} \includegraphics[trim={40 5 40 15},clip,width=0.34\textwidth]{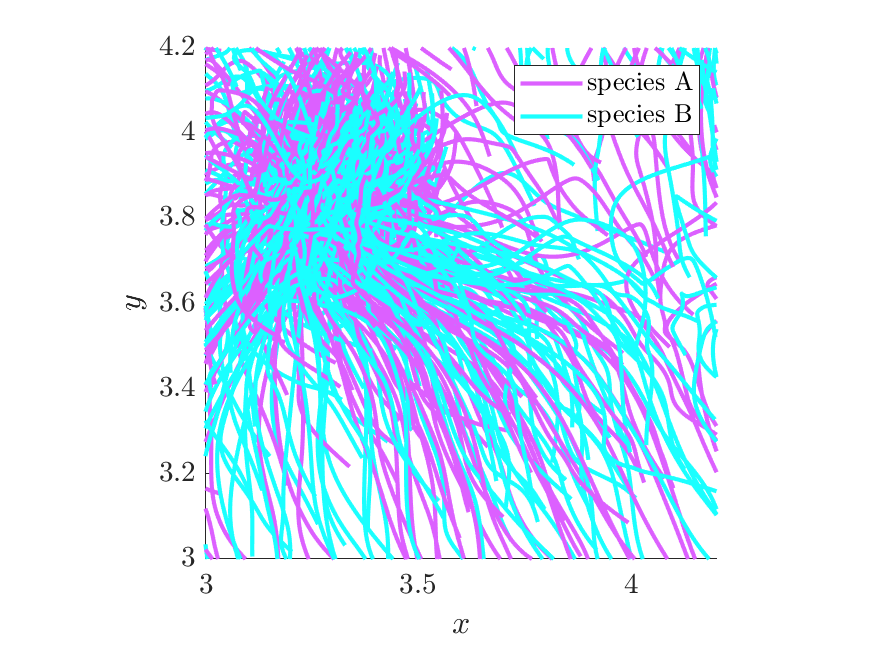} & 		\hspace{-0.9cm} \includegraphics[trim={40 5 40 15},clip,width=0.34\textwidth]{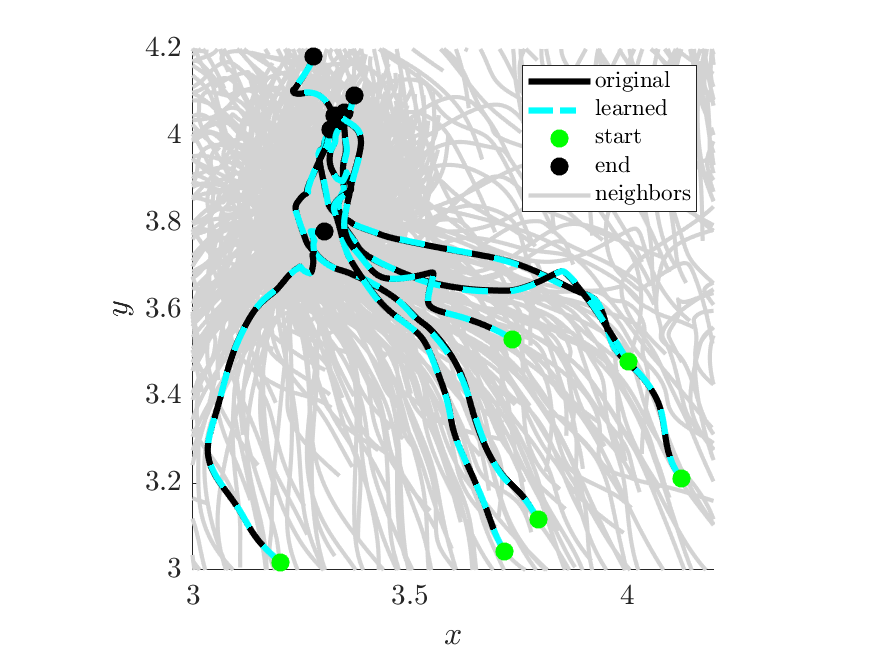} &
\hspace{-0.9cm}    \includegraphics[trim={40 5 40 15},clip,width=0.34\textwidth]{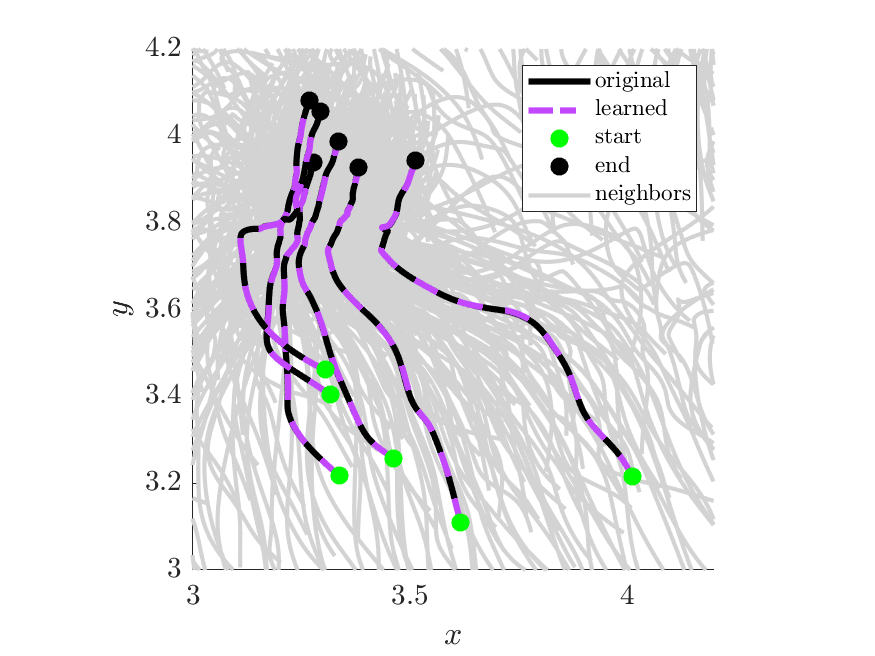} 
\end{tabular}
\caption{Example trajectories from experiment $\Xbf_{A,B}$(long). Cells with true labels are depicted on the left and classified cells from species $B$ (middle) are species $A$ (right) are depicted with model output overlapping the input data in each case.}
\label{trajXAB}
\end{figure*}

Figure \ref{trajXAB} shows a comparison between original and learned trajectories for $\Xbf_{A,B}$(long) and Figure \ref{gmmXAB} depicts representative Gaussian mixture models. In particular, Figure \ref{gmmXAB} (left) shows increased overlap between the two Gaussian mixtures in the first iteration, compared to Figure \ref{gmmXACXBC}, however model performance is still sufficiently different as to classify $\approx 98\%$ of cells correctly.

\subsection{Three-Species Population}

As a final test we identify species from the three-species experiment $\Xbf_{A,B,C}$. Similar to the case $\Xbf_{A,B}$, we see improvements with a longer time-series $\Xbf_{A,B,C}$(long). For the initial experiment $\Xbf_{A,B,C}$, species C is completely identified in the first cluster (Table \ref{tableXABC} row 2), and in the second cluster $98.8\%$ of species $B$ cells are identified along with $9.1\%$ of species $A$ cells, leading to a fairly inaccurate model ($\Delta V\approx 0.06$). The subsequent clusters divide the remaining species $A$ and $B$ cells. 

Doubling the time series with $\Xbf_{A,B,C}$(long), we find the majority of each species residing in its own cluster (Table \ref{tableXABC} row 3). Cluster 1 contains all of the species C cells, cluster 2 consists of 94\% of species B cells and 1.8\% of species A, and cluster 4 consists of 95\% of species A. Moreover, the aggregrate models for each of these clusters result in validation errors under $1\%$.

\begin{figure*}
\begin{tabular}{cc}
	\includegraphics[trim={40 5 40 15},clip,width=0.45\textwidth]{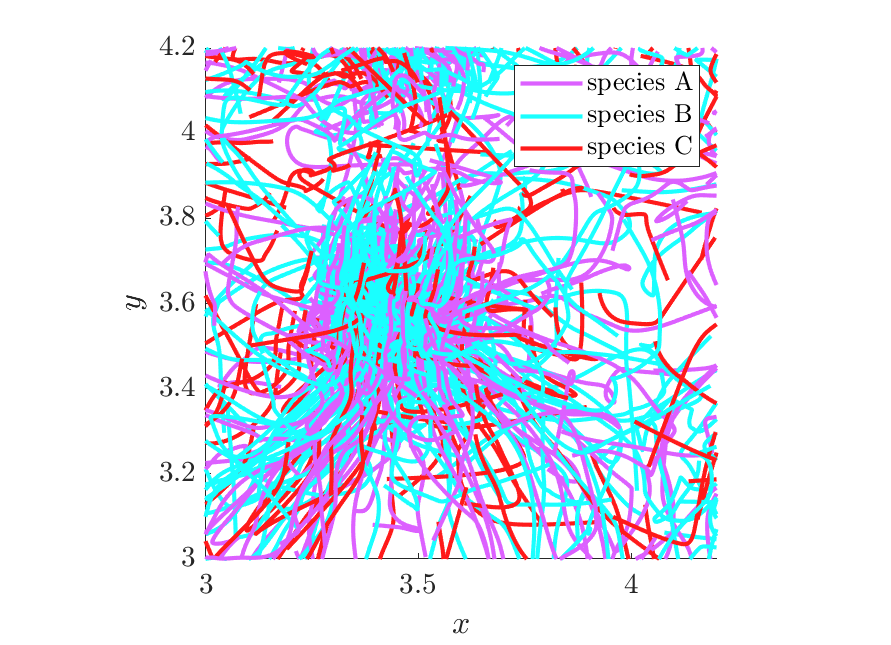} & 		\includegraphics[trim={40 5 40 15},clip,width=0.45\textwidth]{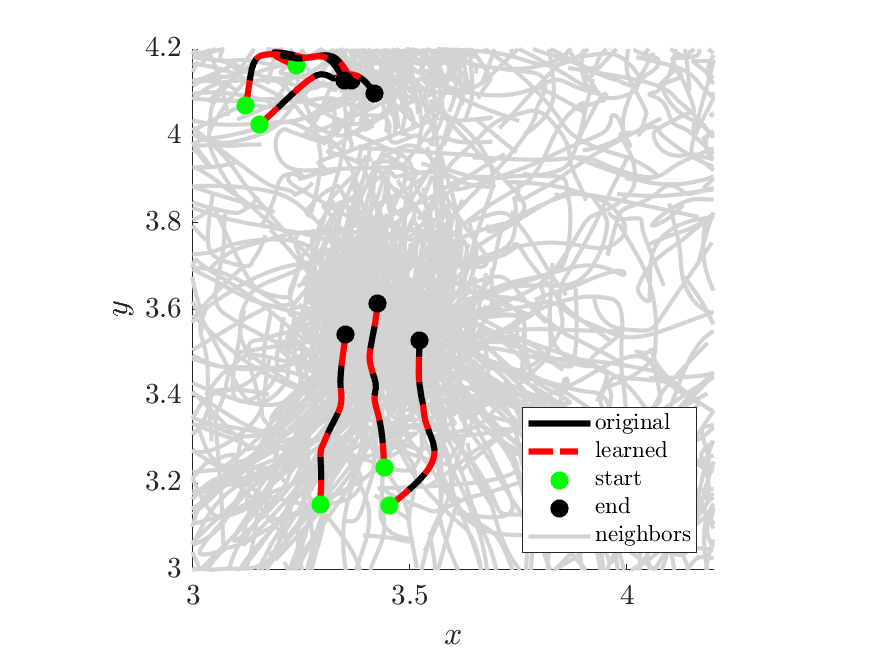} \\	\includegraphics[trim={40 5 40 15},clip,width=0.45\textwidth]{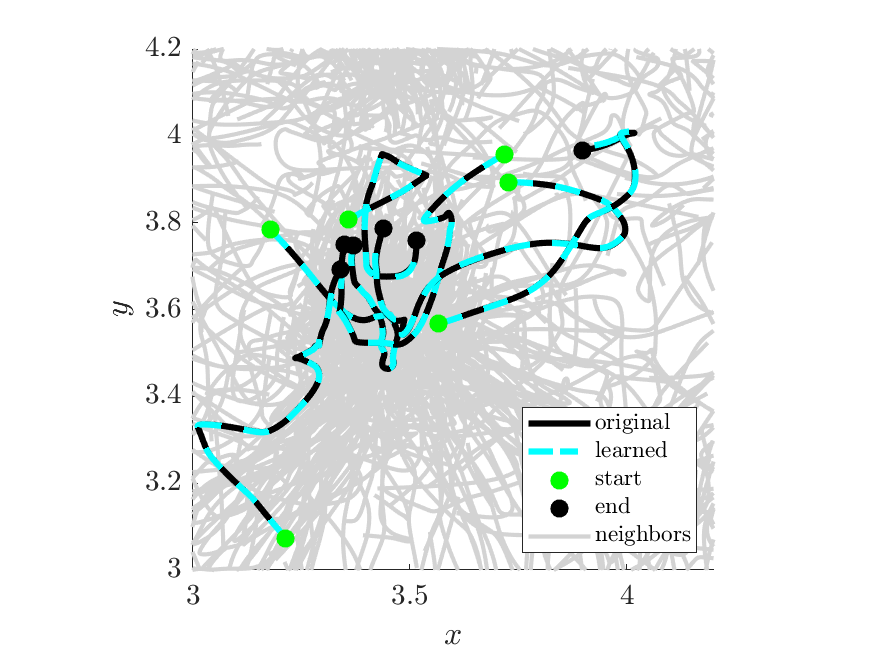} & 		\includegraphics[trim={40 5 40 15},clip,width=0.45\textwidth]{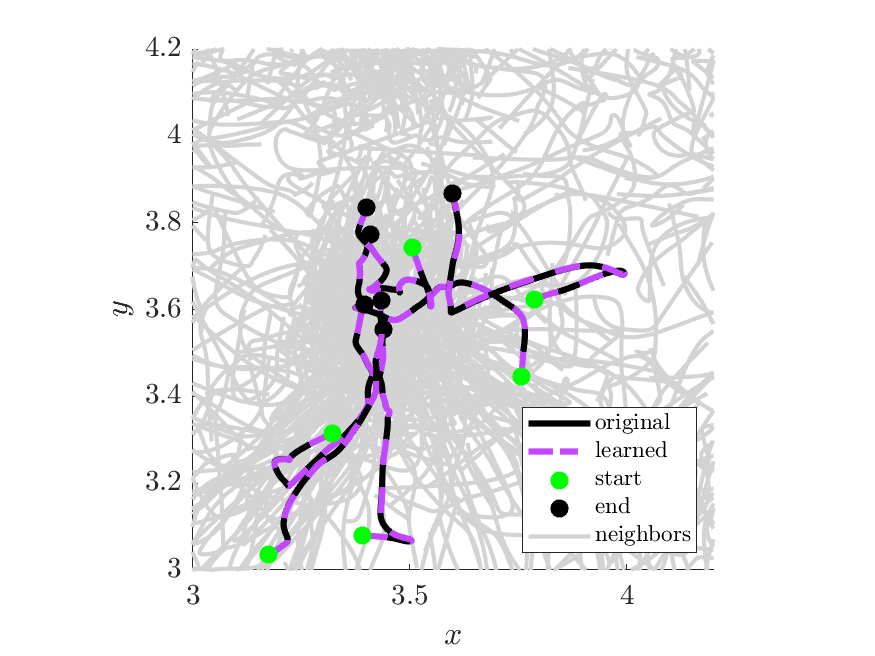}
\end{tabular}
\caption{Example trajectories simulated using learned models from $\Xbf_{A,B,C}$(long). Top left: example cells with true labels. Top right, bottom left, bottom right: example trajectories from clusters 1-3 (see Table \ref{tableXABC} row 3 for details). Note in particular that learned models for clusters 2 and 3 are able to capture sharp turns in the true dynamics.}
\label{trajXABC}
\end{figure*}

Clusters 4 and 5 of $\Xbf_{A,B,C}$(long) are each small mixtures of species A and species B, however the learned forces in each case are nevertheless accurate. We find that $\Delta f\leq 5\%$ (row 3 of Table \ref{tableXABC}, subrows 4 and 5) while the validation error is fairly high, with $\Delta V \geq 6.6\%$. This indicates that the cells in clusters 4 and 5 have trajectories that are particularly sensitive to perturbations. Given the complexity of the dynamics (one can observe sharp turns taken by cells in the bottom two plots of Figure \ref{trajXABC}), trajectories cannot be expected to remain close for all time, and in this case the validation error \eqref{VEformula} may be too strong a metric\footnote{We discuss this further in the conclusion but leave a complete investigation to future work.}. It is remarkable that the aggregate models for clusters 1, 2, and 3 produce accurate learned trajectories for the whole time series.

\begin{table*}
\begin{center}
\begin{tabular}{|c|c|c|c|c|c|r|c|}
\hline
Experiment & $\Delta f_\text{a-r}$ & $\Delta f_\text{align}$ &  $\Delta f_\text{drag}$ & CS($A$)  & CS($B$)  & CS($C$) & $\Delta V$ \\
\hline
$\Xbf_{A,B,C}$ & 
\begin{tabular}{c}
--- \\  $0.0287$ \\ 0.0243 \\ 0.0022 \\0.0478 \end{tabular} & 
\begin{tabular}{c}
0.0005 \\ $\text{---}$ \\ 0.2321  \\ 0.0001 \\ $\text{---}$ \end{tabular} & 
\begin{tabular}{c}
0.0020 \\  0.0191 \\ $\text{---}$ \\ 0.0050  \\ 0.0118 \end{tabular} &
\begin{tabular}{c}
0 \\ 0.091 \\ 0.542  \\ 0.332 \\ 0.009 \end{tabular} & 
\begin{tabular}{c}
0 \\ 0.988 \\ 0.006 \\ 0  \\ 0.006 \end{tabular} &
\begin{tabular}{c}
1.000 \\  0 \\ 0 \\0 \\ 0 \end{tabular} &
\begin{tabular}{c}
0.0016 \\  0.0620 \\ 0.5010 \\ 0.0406  \\ 0.3462 \end{tabular} \\
\hline
$\Xbf_{A,B,C}$(long) & 
\begin{tabular}{c}
--- \\  $0.0117$ \\ 0.0031 \\ 0.0090 \\ 0.0082 \end{tabular} & 
\begin{tabular}{c}
0.0009 \\ --- \\ 0.0034 \\ --- \\ 0.0303 \end{tabular} & 
\begin{tabular}{c}
 0.0043\\  0.0048 \\ 0.0041 \\ 0.0011  \\ 0.0404 \end{tabular} &
\begin{tabular}{c}
0 \\ 0.018  \\ \cellcolor[rgb]{0.7608,0.2784,1.0000}{0.949} \\ 0.003 \\ 0.040 \end{tabular} & 
\begin{tabular}{c}
0 \\ \cellcolor[rgb]{0,1,1}{0.940} \\ 0 \\ 0.06  \\ 0 \end{tabular} &
\begin{tabular}{c}
\cellcolor[rgb]{1,0,0}{1.000}\\  0 \\ 0 \\0 \\0 \end{tabular} &
\begin{tabular}{c}
0.0011 \\  0.0095 \\ 0.0094 \\ 0.0663  \\ 0.0951 \end{tabular} \\ \hline
\end{tabular}
\end{center}
\caption{Performance of model learning and classification algorithm of three-species populations. $\Xbf_{A,B,C}$(long) is simply the continuation of $\Xbf_{A,B,C}$ to twice the time horizon.}
\label{tableXABC}
\end{table*}

\section{Discussion}\label{sec:discussion}

We have introduced a method for performing a combined classification and model selection task relevant to heterogeneous systems of autonomous agents. Specifically, we have shown that learning an ensemble of interacting particle models (one for each agent) allows iterative classification of agents into species according to their forward simulation accuracy. This is surprising due to the limited information carried in a single trajectory. Fortunately, the validation errors empirically approximate a log-normal distribution, hence Gaussian mixture model classification arises as the appropriate tool for identifying species membership.
    
Computational feasibility of this approach is grounded in the parallel nature of both the learning algorithm and the simulation component. Learning each single-trajectory model is cheap, with linear systems of size $m\times n$ with $m$ and $n$ not exceeding several hundreds (see Table \ref{table:time} for walltimes). In the simulation step, each trajectory is validated separately and in parallel, and neighbor interactions are computed using the measurement data itself, resulting in $\CalO(N)$ pairwise interaction computations per timestep instead of an $\CalO(N^2)$ full simulation\footnote{We are not aware of other simulation approaches in the context of interacting particles that use the measurement data for direct calculation of the forces.}.

Several aspects of this approach deserve a more technical analysis, which may lead to improvements. We aim in future work to undertake more rigorous study of each component of the algorithm as outlined below.

\subsubsection*{Learning single-cell models} 

\begin{enumerate}[label=(\Roman*)]
    \item {\bf Information content}: It would be beneficial to quantify the information content in each cell trajectory, possibly eliminating trajectories that do not provide sufficient information. We saw in the experiments $\Xbf_{A,B}$ and $\Xbf_{A,B,C}$ that increasing the length of the trajectory leads to better classification when two species exhibit similar dynamics. It may be possible to use existing techniques, such as force matching, to identify highly-informative cells from forces magnitudes, neighbor distributions, etc. Model replacement, as outlined in Appendix \ref{app:modelrep}, is an initial step in this direction.\\
    
    \item {\bf Noisy trajectories}: While we expect measurement noise to be filtered out by the weak form to similar extent demonstrated on ODE systems \cite{messenger2020weak}, we leave examination of the robustness to both intrinsic (e.g.\ Brownian) and extrinsic (e.g.\ measurement) noise to future work.\\
    
    \item {\bf Model library}: We chose the force bases and constraints to reflect physical properties, namely short-distance repulsion, long-distance decay, negative alignment, and negative drag. Directional modes enforce bilateral symmetry, and are low order (monopole, dipole, quadrupole). These can easily be adapted to incorporate other known information, moreover the bases themselves may be adapted to the data (note this is partially done, using the neighbor distance distribution $\rho_{rr}$ to restrict the range of interactions). One major assumption here is that there is no propulsion force, that energy is increased only through anisotropic interactions with neighbors. It would be interesting to examine whether this assumption holds true.\\
    
    
    \item {\bf Regression approach}: We employ modified sequential thresholding, which looks for an overall sparse solution, although we threshold {\it only} on the term magnitude $\nrm{\Gbf_j\wbf_j}$ and not that raw coefficient $\wbf_j$. This in particular allows $f_\text{a-r}, f_\text{align}, f_\text{drag}$ to have equal opportunity to enter the model despite different scales and bases used. The effect is that the resulting model is sparse in the {\it force modes} (as described in Section \ref{step:cluster}), while each force mode may have many components (in fact $f_\text{a-r}$ is usually not sparse on a given directional mode). It may be more appropriate to use a group sparsity-enforcing method, such as constrained group LASSO. In general, this lies at the intersection of approximation and selection, where sparse selection is required to select the correct modes, however the force content in each mode requires approximation. Explorations of the appropriate balance between selection and approximation would be valuable.
    
    
\end{enumerate}

\subsubsection*{Cluster \& Aggregate}

\begin{enumerate}[label=(\Roman*)]
 \item Here we cluster models based on the directional modes present. This can easily be extended to the full pattern of nonzero elements in the model vector $\what$, although this depends on the number of resulting clusters.  
\item We have used a simple uniformly-weighted average \eqref{wbar} to aggregate models, however the use of model replacement (see Appendix \ref{app:modelrep}) implies that the average is implicitly weighted according to {\it model generalizability}. Results may be improved if other criteria (e.g.\ information criteria) are incorporated into the weighted average, or if the median is taken instead of the mean.
\item In the examples above, the aggregate model is used as the final model for the give class. Instead, one could further refine the model by performing an additional regression combining all data from the identified species.
\end{enumerate}

\subsubsection*{Validate \& Classify} 

In practice, the validation errors can easily be checked to satisfy log-normalcy {\it a posteriori}. If this is not satisfied, it may not be straightforward to cluster based on the validation error. In particular, chaotic trajectories cannot be expected to achieve a low validation error, in which case another metric is needed. In this case, it is reasonable to require that trajectories to be long enough to compute statistics. We aim to investigate the requirements for performing classification with chaotic interacting particles in future work.

\appendix

\section{Notation, Algorithm Hyperparameters, and Walltimes}\label{app:hp}

\begin{table*}
\begin{center}
\begin{tabular}{|c|c|}
\hline Symbol & Definition \\ \hline
$f_\text{a-r}$ & attractive-repulsive force \\ \hline
$f_\text{align}$ & alignment force \\ \hline
$f_\text{drag}$ & drag force \\ \hline
$(x_i,v_i)$ & position and velocity of cell $i$ \\ \hline
$\theta_{ij}$ & angle between $v_i$ and $x_i-x_j$ \\ \hline
$N$ & number of focal cells selected for learning  \\ \hline
$N_\text{tot}$ & total number of cells in the population  \\ \hline
$\eta_i$ & noise in force law for cell $i$  \\ \hline
$r_{nf}$ & near-field radius, below which $f_\text{a-r}$ is repulsive  \\ \hline
$r_{ff}$ & far-field radius, above which $f_\text{a-r}$ is attractive   \\ \hline
$F_i$ & total force on particle $i$  \\ \hline
$(X,V)$ & entire particle system position and velocity  \\ \hline
$(\Xbf,\Vbf)$ & particle position and velocity time-series data  \\ \hline
$\Theta(X,V)=(f_j(X,V))_{1\leq j \leq J}$ & basis of force functions for learning  \\ \hline
$(\Theta(\Xbf,\Vbf),\ddot{\Xbf})$ & linear system for learning using SINDy  \\ \hline
$(\Gbf,\bbf)$ & linear system for learning using WSINDy  \\ \hline
$\wstar$ & true model coefficients  \\ \hline
$\{\phi_q\}_{q\in Q}$ & test functions to compute weak time derivatives  \\ \hline
$(m,p,t_q)$ & test function hyperparameters  \\ \hline
$\CalF_\text{a-r}$, $\CalF_\text{align}$, and $\CalF_\text{drag}$ & trial bases  \\ \hline
$p_\ell$ & Laguerre polynomials  \\ \hline
$r_{\max}$ & maximum inter-particle distance observed in $\Xbf$  \\ \hline
$\CalM$ & set of single-cell models  \\ \hline
$f_\text{force}^{(i)}$ & directional force modes (force $\in \{\text{a-r},\text{align},\text{drag}\}$) \\ \hline
$\CalC$ & set of model clusters based on force modes  \\ \hline
$\overline{\CalC}$ & cluster with the most members  \\ \hline
$\overline{\wbf}$ & coefficients obtained from averaging the models in cluster $\overline{\CalC}$  \\ \hline
$\overline{\CalM}$ & model associated with coefficients $\overline{\wbf}$  \\ \hline
$\overline{\CalS}$ & species identified as obeying the model $\overline{\CalM}$  \\ \hline
$S$ & Number of species identified  \\ \hline
$(\overline{x}_i,\overline{v}_i)$ & simulated cells using $\overline{\CalM}$  \\ \hline
$L$ & number of time steps in data  \\ \hline
$\Delta t$ & time step of data  \\ \hline
$\Delta t'$ & time step for simulation  \\ \hline
$\overline{\CalM}(x,v,X,V)$ & force on $(x,v)$ resulting from neighbors $(X,V)$ using model $\overline{\CalM}$ \\\hline
$(\Xbf^{\prime i},\Vbf^{\prime i})$ & set of particle positions and velocities with $i$th cell removed  \\ \hline
$\Delta V_i$ & validation error of cell $i$  \\ \hline
$VE$ & set of validation errors  \\ \hline
\end{tabular}
\end{center}
\caption{Summary of notation used througout.}\label{table:notation}
\end{table*}

In Table \ref{table:notation} we include all notation used throughout the article. In Table \ref{tabel:hp} we list several hyperparameter choices for the examples above which were not covered in Sections \ref{sec:alg}-\ref{sec:results}. Finally, we include walltimes for the main components of the algorithm in Table \ref{table:time}, recorded in MATLAB using an AMD Ryzen 7 pro 4750u processor.

\begin{table*}
\begin{center}
\begin{tabular}{|c|c|c|c|c|}
\hline Experiment & $m$ & $p$ & $r_{nf}$ & $r_{ff}$ \\\hline
$\Xbf_A$ & $38$ & $8$ & $0.0497$ & 1\\\hline
$\Xbf_B$ & $38$ & $8$ & $0.0365$ & 1\\\hline
$\Xbf_C$ & $32$ & $9$ & $0.0219$ & 1\\\hline
$\Xbf_{A,C}$ & $35$ & $9$ & $0.0494$ & 1\\\hline
$\Xbf_{B,C}$ & $35$ & $9$ & $0.0499$ & 1\\\hline
$\Xbf_{A,B}$ & $38$ & $8$ & $0.0365$ & 1\\\hline
$\Xbf_{A,B}$(long) & $31$ & $9$ & $0.0219$ & 1\\\hline
$\Xbf_{A,B,C}$ & $38$ & $8$ & $0.0569$ & 1\\\hline
$\Xbf_{A,B,C}$(long) & $31$ & $9$ & $0.0253$ & 1\\\hline
\end{tabular}
\end{center}
\caption{Hyperparameters used for each example.}
\label{tabel:hp}
\end{table*}

\begin{table*}
\begin{center}
\begin{tabular}{|c|c|c|c|}
\hline $\xbf_i\to\CalM_i$ & $\CalM\to (\overline{\CalC},\overline{\CalM})$ & $(\xbf_i,\overline{\CalM})\to \Delta V_i$ & $VE\to \overline{\CalS}$ \\
\hline 5-10 sec. &  10-30 sec. & 10-30 sec. &  $<1$ sec.\\  
\hline
\end{tabular}
\end{center}
\caption{Wall times for main components of one iteration of the algorithm, recorded for the $\Xbf_{A,B,C}$(long) experiment ($N_\text{tot}=1000$ cells and $L=400$ timepoints). Each component contains an inner iteration which may be trivially parallelized. The reported times are for one step of the respective inner iteration. Specifically, it takes 5-10 seconds to learn each model $\CalM_i$, while computation time for the full set of models $\CalM = \{\CalM_1,\dots,\CalM_N\}$ depends on the number of available CPUs. To form the model clusters $\CalC=\{\CalC_1,\dots\CalC_r\}$, find the largest cluster $\overline{\CalC}$, and compute the averaged model $\overline{\CalM}$, forward simulations of each model $\CalM_i$ (see Section \ref{app:modelrep}) are performed which take 10-30 seconds (depending on the complexity of $\CalM_i$). Similarly, computation of each $\Delta V_i$ requires one forward simulation (10-30 seconds). Lastly, to identify $\overline{\CalS}$, it takes $<1$ second to perform GMM classification of $\log_{10}(VE)$, and the results of 20 rounds of classification are averaged. If full parallelization is available, the walltime is less than two minutes per outer iteration, or $<2S$ minutes in total, where $S$ is the number of identified species. For comparison, the cost of generating the data for $\Xbf_{A,B,C}$(long) takes 5-6 hours.}
\label{table:time}
\end{table*}

\section{Simulation details}

All example data was generated from the exact models using Forward Euler with a timestep of $\Delta t^* \approx 0.00042$ up until a final time of $T\approx 26$. The time series was then coarsened to a resolution of $\Delta t=0.13$, resulting in a total of $L = 200$ time points for learning. Experiments $\Xbf_{A,B}$(long) and $\Xbf_{A,B,C}$(long) are simply extensions of $\Xbf_{A,B}$ or $\Xbf_{A,B,C}$ by an addition 200 timepoints at the same resolution. Initial positions were generated using Latin hypercube sampling in a box of side length $2$, while initial velocities were drawn from a Gaussian distribution with mean 0 and covariance $0.0025 \Ibf_2$.
 


\section{Constrained sparse regression}\label{app:csr}

The constrained sequential thresholding algorithm requires solving at each thresholding iteration $\ell$ a linearly constrained quadratic program of the form
\begin{equation}
\wbf^{(\ell+1)} = \argmin_{\substack{\wbf \text{ s.t. }\Cbf\wbf \leq \dbf \\ \supp{\wbf}\subset \CalI^{(\ell)}}} \nrm{\Gbf\wbf-\bbf}_2^2
\end{equation}
where $\CalI^{(\ell)}$ is the set of coefficients of $\wbf^{(\ell)}$ satisfying \eqref{thresh}. The constraint system $\Cbf\wbf \leq \dbf$ has the following four components. 
\begin{enumerate}
\item \underline{$f_\text{a-r}\geq 0$ when $0\leq r< r_{nf}$}: for the near-field repulsion of $f_\text{a-r}$ we discretize the region $\{(r,\theta)\ :\ 0\leq r< r_{nf}, \ \theta\in [0,2\pi)\}$ choosing 5 equally-spaced points in $r$ from $10^{-6}$ to $r_{nf}$ and 5 equally-spaced points in $\theta$ from $0$ to $\pi$. Evaluating each of the basis function for $f_\text{a-r}$ at this grid results in a constraint system $\Cbf_{\text{a-r},nf}\wbf_\text{a-r}\leq \textbf{0}$ of dimension $25\times J_\text{a-r}$ where $J_\text{a-r}$ is the number of basis functions used to approximate $f_\text{a-r}$ and $\wbf_\text{a-r}$ is the restriction of $\wbf$ to coefficients of $f_\text{a-r}$.
\item \underline{$f_\text{a-r}\leq 0$ when $r\geq  r_{ff}$}: similarly for the far-field region, we choose $10$ equally-spaced points in $r$ from $r_{ff}$ to $r_{\max}$, where $r_{\max}$ is the maximum observed neighbor-neighbor distance in the simulation, and $\theta$ over the same points as previous. This results in a constraint system $\Cbf_{\text{a-r},ff}\wbf_\text{a-r}\leq \textbf{0}$ of dimensions $50\times J_\text{a-r}$.
\item \underline{$f_\text{align}\leq 0$}: since the basis is positive, the constrain system here is simply $\Ibf_{J_\text{align}}\wbf_\text{align}\leq \textbf{0}$, where $\Ibf_n$ indicates the identity on $\Rbb^n$. 
\item \underline{$f_\text{drag}\leq 0$}: similarly the basis is positive, so the constraint system is  
$\Ibf_{J_\text{drag}}\wbf_\text{drag}\leq \textbf{0}$.
\end{enumerate}
Altogether we get $\dbf = \textbf{0}$ and
\[\Cbf = \begin{bmatrix} 
\Cbf_{\text{a-r},nf} & 0 & 0 \\
\Cbf_{\text{a-r},ff} & 0 & 0 \\
0 & \Ibf_{J_\text{align}} & 0 \\
0 & 0 & \Ibf_{J_\text{drag}} \end{bmatrix}\]

We use MATLAB's \texttt{quadprog} with constraint tolerance $10^{-10}$ and maximum iterations set to 1000. Note that since $\dbf = \textbf{0}$, we do not lose feasibility during the thresholding step, which is possible in general.

\section{Model replacement}\label{app:modelrep}

Once the initial batch of $N$ models $\CalM$ is learned, we simulate each model $\CalM_i$ as outlined in \ref{sec:validate} on $K$ different {\it validation cells} selected from the data, where $K=20$ throughout. For a given model $\CalM_i$, we select these $K$ validation cells by finding cells in the population that match well certain statistics of cell $i$. In particular, we define the following distributions:
\begin{align*}
\rho^{(i)}_{rr}(r) &= \frac{1}{T}\int_0^T\Pbb_{x\sim X^\prime}\left(\| x_i(t)-x(t)\|<r\right)dt\\
\rho^{(i)}_{vv}(s) &= \frac{1}{T}\int_0^T\Pbb_{v\sim V^\prime}\left(\| v_i(t)-v(t)\|<s\right)dt\\
\rho^{(i)}_{v}(s) &= \Pbb \left(\|v_i\|<s\right)
\end{align*}
where $(X^\prime,V^\prime)$ denote the remainder of the cell population excluding cell $i$. Respectively, these denote the distribution of {\it distances} from cell $i$ to all other cells, the distribution of {\it velocity differences} between cell $i$ and all other cells, and the distribution of {\it speeds} that cell $i$ experiences\footnote{These statistics are likely to correspond to the information content that cell $i$ carries about its own forces $f_\text{a-r}$, $f_\text{align}$, and $f_\text{drag}$, given the force dependencies}. We approximate these distributions from the data using histograms with 50 bins.

For each cell $i$ we compute the Kullback-Leibler (KL) divergence between its distributions $\rho^{(i)}_{rr}$, $\rho^{(i)}_{vv}$, $\rho^{(i)}_{v}$ and those of the rest of the population\footnote{To limit the computational overhead, we only compute these KL divergences to cell $i$'s nearest 200 neighbors in the Euclidean sense.}, where the KL divergence between densities $\rho$ and $\nu$ is given by 
\[\CalD_{KL}(\rho|\nu) = -\int \rho(x) \log \left(\frac{\nu(x)}{\rho(x)}\right)\,dx.\]
The $K$ validation cells used to validate model $i$ are the $K$ cells with smallest cost $\scL$, defined by 
\[\scL:=\CalD_{KL}(\rho^{(i)}_{rr}|\rho^{(j)}_{rr})^2+\CalD_{KL}(\rho^{(i)}_{vv}|\rho^{(j)}_{vv})^2+\CalD_{KL}(\rho^{(i)}_{v}|\rho^{(j)}_{v})^2.\]

Let the validation error $\Delta V_{i\to j}$ be defined as in \eqref{VEformula}, but indicating $\CalM_i$ used to validate cell $j$ (i.e.\ using the initial conditions of cell $j$). We replace $\CalM_i$ with $\CalM_j$ if the following three conditions are met:

\begin{enumerate}[label=(\arabic*)]
\item $\Delta V_{i\to i} > \Delta V_{j\to i}$
\item $\Delta V_{i\to j} > \Delta V_{j\to j}$
\item $\max\{\Delta V_{j\to i},\Delta V_{j\to j}\}<\text{tol}$
\end{enumerate}
where we set $\text{tol}=0.25$ in this work. In words, $\CalM_j$ replaces $\CalM_i$ if (1) $\CalM_j$  performs better than $\CalM_i$ on cell $i$, (2) $\CalM_j$ performs better than $\CalM_i$ on cell $j$, and (3) $\CalM_j$ achieves a reasonably low error (defined by tol) on both cell $i$ and cell $j$. (Note that cell $i$ and cell $j$ are required to be mutual validation cells for a model replacement to occur). Furthermore, if $\CalM_j$ replaces $\CalM_i$, and another model $\CalM_k$ replaces $\CalM_j$, we reassign $\CalM_i$ to $\CalM_k$ as well, even if cells $i$ and $k$ are not mutual validation cells.  

We find this approach to be crucial to increasing accuracy of the learned models, as it transfers successful learning of few cells with highly informative trajectories to cells with less informative trajectories. As with all validation steps of our algorithm, this approach would be infeasible if not for fast data-driven forward simulations. 

\section{Gaussian Mixture Model (GMM) classification}\label{app:GMM}

Since the GMM fitting is performed using the expectation-maximization algorithm with random conditions, we perform the GMM fitting for 20 trials and identify $\overline{\CalS}$ as the cells that in more than half of the trials appear in the mixture with lowest error.

At some stage in the algorithm, all the remaining cells will be homogeneous. In this case, a 2-species Gaussian is the wrong model. To account for this, we do an initial fit to a single Gaussian and compute its Bayesian Information Criterion (BIC). We accept the 2-mixture GMM if the average BIC of all 20 trial GMM fits with 2 mixtures is lower than that of the single Gaussian.

\end{multicols}

\subsubsection*{Data Access} All software used to generate the results in this work is available at this repository: \url{https://github.com/MathBioCU/WSINDy_CellCluster.git}

\subsubsection*{Funding} This research was supported in part by the NSF/NIH Joint DMS/NIGMS Mathematical Biology Initiative grant R01GM126559, in part by the NSF Mathematical Biology MODULUS grant 2054085, and in part by the NSF Computing and Communications Foundations grant 1815983. This work also utilized resources from the University of Colorado Boulder Research Computing Group, which is supported by the National Science Foundation (awards ACI-1532235 and ACI-1532236), the University of Colorado Boulder, and Colorado
State University.

\subsubsection*{Acknowledgements} The authors wish to thank Prof.\,Vanja Duki\'{c} (Department of Applied Mathematics, University of Colorado, Boulder) for insightful comments about statistical aspects of this work as well as the mathematical form of the directional interaction kernel.

\bibliographystyle{plain}
\bibliography{refs}

\end{document}

%% file: ex_shared.tex
\newsiamremark{remark}{Remark}
\newsiamremark{hypothesis}{Hypothesis}
\crefname{hypothesis}{Hypothesis}{Hypotheses}
\newsiamthm{claim}{Claim}

\headers{Learning Heterogeneous Agent Populations}{D. Messenger, G. Wheeler, X. Liu, D. Bortz.}

\title{Learning Anisotropic Interaction Rules from Individual Trajectories in a Heterogeneous Cellular Population}

\author{Daniel A. Messenger\thanks{Department of Applied Mathematics, University of Colorado, Boulder, CO 80309-0526, USA. 
	(\email{daniel.messenger@colorado.edu}, \email{david.bortz@colorado.edu}).}
\and Graycen E. Wheeler \thanks{Department of Biochemistry, University of Colorado, Boulder, CO 80309-0526, USA.
	(\email{graycen.wheeler@colorado.edu}, \email{Xuedong.Liu@colorado.edu}).}
\and Xuedong Liu\footnotemark[2]
\and David M. Bortz\footnotemark[1]}

\usepackage{amsopn}

\usepackage{lipsum}
\usepackage{amsfonts}
\usepackage{graphicx}
\usepackage{epstopdf}
\usepackage{subcaption}
\usepackage{algorithmic}
\usepackage{multicol}
\usepackage{tikz,empheq,mathrsfs,colortbl}

\ifpdf
  \DeclareGraphicsExtensions{.eps,.pdf,.png,.jpg}
\else
  \DeclareGraphicsExtensions{.eps}
\fi

\usepackage[latin9]{inputenc}
\usepackage{geometry}
\usepackage{xcolor}
\usepackage{pdfcolmk}
\usepackage{float}
\usepackage{mathtools}
\usepackage{amsbsy}
\usepackage{amsmath}
\usepackage{amstext}
\usepackage{amssymb}
\usepackage{graphicx}
\usepackage{enumitem}
\PassOptionsToPackage{normalem}{ulem}
\usepackage{ulem}

\graphicspath{{figures/}}

\makeatletter

\providecolor{lyxadded}{rgb}{0,0,1}
\providecolor{lyxdeleted}{rgb}{1,0,0}

\DeclareRobustCommand{\lyxsout}[1]{\ifx\\#1\else\sout{#1}\fi}

\numberwithin{equation}{section}
\numberwithin{figure}{section}
\theoremstyle{plain}

  \theoremstyle{remark}
  
  \theoremstyle{plain}

\input{preamble}

\input{notation}

\usepackage{stmaryrd}

\makeatother

  \providecommand{\lemmaname}{Lemma}
  \providecommand{\remarkname}{Remark}
\providecommand{\theoremname}{Theorem}


%% file: preamble.tex
\usepackage{
amsmath,
amssymb,
appendix,
bm,
bbm,
caption,
color,
epsf,
enumitem,
float,
graphicx,
hyperref,
listings,
mathtools,
mathrsfs,
subcaption,
tikz,
titletoc,
url,
ulem,
xcolor,
mathtools
}

\DeclareMathOperator*{\argmin}{arg\,min}

\hypersetup{
    colorlinks=true, 
    linktoc=all,     
    linkcolor=black,  
}

\numberwithin{equation}{section}

\tikzset{every label/.style={font=\footnotesize,inner sep=1pt}}
\DeclareCaptionFormat{myformat}{#1#2#3\hrulefill}
\graphicspath{{figures/}}

%% file: notation.tex
\newcommand{\Pbb}{\mathbb{P}}

\newcommand{\Rbb}{\mathbb{R}}

\newcommand{\bbf}{\mathbf{b}}

\newcommand{\Cbf}{\mathbf{C}}
\newcommand{\dbf}{\mathbf{d}}

\newcommand{\Gbf}{\mathbf{G}}

\newcommand{\Ibf}{\mathbf{I}}

\newcommand{\vbf}{\mathbf{v}}
\newcommand{\Vbf}{\mathbf{V}}
\newcommand{\wbf}{\mathbf{w}}

\newcommand{\xbf}{\mathbf{x}}
\newcommand{\Xbf}{\mathbf{X}}

\newcommand{\ep}{\epsilon}

\newcommand{\CalC}{{\mathcal{C}}}
\newcommand{\CalD}{{\mathcal{D}}}

\newcommand{\CalF}{{\mathcal{F}}}

\newcommand{\CalI}{{\mathcal{I}}}

\newcommand{\CalM}{{\mathcal{M}}}

\newcommand{\CalO}{{\mathcal{O}}}

\newcommand{\CalS}{{\mathcal{S}}}

\newcommand{\scL}{\mathscr{L}}

\newcommand{\nrm}[1]{\left\Vert {#1} \right\Vert}

\newcommand{\supp}[1]{\text{supp}\left(#1\right)}

\renewcommand{\tilde}[1]{\widetilde{#1}}

\newcommand{\lan}{\left\langle}
\newcommand{\ran}{\right\rangle}

\newcommand{\wstar}{\wbf^\star}
\newcommand{\what}{{\widehat{\wbf}}}

\theoremstyle{definition}

%% file: tikz_theta.tex
\begin{center}
\tikzset{every picture/.style={line width=0.75pt}} 
\begin{tikzpicture}[x=0.75pt,y=0.75pt,yscale=-1,xscale=1]
\draw  [fill={rgb, 255:red, 65; green, 117; blue, 5 }  ,fill opacity=1 ] (225,219.19) .. controls (225,211.56) and (231.63,205.38) .. (239.8,205.38) .. controls (247.97,205.38) and (254.6,211.56) .. (254.6,219.19) .. controls (254.6,226.82) and (247.97,233) .. (239.8,233) .. controls (231.63,233) and (225,226.82) .. (225,219.19) -- cycle ;
\draw  [fill={rgb, 255:red, 208; green, 2; blue, 27 }  ,fill opacity=1 ] (405.4,151.59) .. controls (405.4,143.97) and (412.03,137.78) .. (420.2,137.78) .. controls (428.37,137.78) and (435,143.97) .. (435,151.59) .. controls (435,159.22) and (428.37,165.4) .. (420.2,165.4) .. controls (412.03,165.4) and (405.4,159.22) .. (405.4,151.59) -- cycle ;
\draw [color={rgb, 255:red, 65; green, 117; blue, 5 }  ,draw opacity=1 ]   (244.45,205.38) -- (262.57,142.05) -- (276.29,96.56) ;
\draw [shift={(276.87,94.65)}, rotate = 106.78] [color={rgb, 255:red, 65; green, 117; blue, 5 }  ,draw opacity=1 ][line width=0.75]    (21.86,-6.58) .. controls (13.9,-2.79) and (6.61,-0.6) .. (0,0) .. controls (6.61,0.6) and (13.9,2.79) .. (21.86,6.58)   ;
\draw  [color={rgb, 255:red, 0; green, 0; blue, 0 }  ,draw opacity=1 ][line width=0.75] [line join = round][line cap = round] (250.21,185.77) .. controls (266.82,184.38) and (271.7,197.47) .. (277.02,204.81) ;
\draw [color={rgb, 255:red, 208; green, 2; blue, 27 }  ,draw opacity=1 ]   (416.11,137.26) -- (400.07,84.72) ;
\draw [shift={(399.48,82.81)}, rotate = 73.01] [color={rgb, 255:red, 208; green, 2; blue, 27 }  ,draw opacity=1 ][line width=0.75]    (10.93,-3.29) .. controls (6.95,-1.4) and (3.31,-0.3) .. (0,0) .. controls (3.31,0.3) and (6.95,1.4) .. (10.93,3.29)   ;
\draw    (254.03,213.8) -- (406.05,160.8) ;
\draw [shift={(407.94,160.14)}, rotate = 160.78] [color={rgb, 255:red, 0; green, 0; blue, 0 }  ][line width=0.75]    (10.93,-3.29) .. controls (6.95,-1.4) and (3.31,-0.3) .. (0,0) .. controls (3.31,0.3) and (6.95,1.4) .. (10.93,3.29)   ;

\draw (274.81,165.34) node [anchor=north west][inner sep=0.75pt]   [align=left] {$\displaystyle \theta _{i}{}_{j}$};
\draw (231.59,207.69) node [anchor=north west][inner sep=0.75pt]   [align=left] {$\displaystyle x_{i}$};
\draw (412.82,141.02) node [anchor=north west][inner sep=0.75pt]   [align=left] {$\displaystyle x_{j}$};
\draw (273.27,75.56) node [anchor=north west][inner sep=0.75pt]   [align=left] {$\displaystyle v_{i}$};
\draw (386.88,62.69) node [anchor=north west][inner sep=0.75pt]   [align=left] {$\displaystyle v_{j}$};
\draw (318.44,190.3) node [anchor=north west][inner sep=0.75pt]  [rotate=-341.27] [align=left] {$\displaystyle x_{j} -x_{i}$};

\end{tikzpicture}
\end{center}